\documentclass[aps,superscriptaddress,notitlepage,nofootinbib,pra,twocolumn]{revtex4-1}

\usepackage{graphicx}
\usepackage{blindtext}
\usepackage[utf8]{inputenc}
\usepackage[space]{grffile}
\usepackage{latexsym}
\usepackage{textcomp}
\usepackage{multirow,booktabs}
\usepackage{amsfonts}
\usepackage{url}
\usepackage{hyperref}
\hypersetup{colorlinks=false,pdfborder={0 0 0}}
\usepackage[utf8]{inputenc}
\usepackage[english]{babel}
\usepackage{graphicx,amsmath,amssymb,color,array}
\usepackage[normalem]{ulem}
\usepackage{multirow}
\usepackage{times}
\usepackage{amsthm}
\usepackage{mathrsfs}
\usepackage{natbib}
\usepackage{color,verbatim,graphics}
\usepackage{psfrag}
\usepackage{tcolorbox}
\usepackage{soul}

\newcommand{\ba}{\begin{eqnarray}}
\newcommand{\be}{\begin{equation}}
\newcommand{\ee}{\end{equation}}

\newcommand{\bn}{\begin{equation*}}
\newcommand{\en}{\end{equation*}}
\newcommand{\ea}{\end{eqnarray}}
\newcommand{\ban}{\begin{eqnarray*}}
\newcommand{\ean}{\end{eqnarray*}}
\newcommand{\Tr}{\operatorname{tr}}
\newcommand{\tr}{\operatorname{tr}}

\definecolor{ivan}{rgb}{0.7,0,0.7}

\newcommand{\sket}[1]{{\ensuremath{\lvert#1\rangle}}}
\newcommand{\lket}[1]{{\ensuremath{\left\lvert#1\right\rangle}}}
\newcommand{\ket}[1]{\if@display\lket{#1}\else\sket{#1}\fi}
\newcommand{\sbra}[1]{{\ensuremath{\langle#1\rvert}}}
\newcommand{\lbra}[1]{{\ensuremath{\left\langle#1\right\rvert}}}
\newcommand{\bra}[1]{\if@display\lbra{#1}\else\sbra{#1}\fi}
\newcommand{\sbraket}[2]{{\ensuremath{\langle#1\rvert#2\rangle}}}
\newcommand{\lbraket}[2]{{\ensuremath{\left\langle#1\!\left\rvert\vphantom{#1}#2\right.\!\right\rangle}}}
\newcommand{\braket}[2]{\if@display\lbraket{#1}{#2}\else\sbraket{#1}{#2}\fi}

\newcommand{\sketbra}[2]{{\ensuremath{\lvert #1\rangle\!\langle #2\rvert}}}
\newcommand{\lketbra}[2]{{\ensuremath{\left\lvert #1\right\rangle\!\!\left\langle #2\right\rvert}}}
\newcommand{\ketbra}[2]{\if@display\lketbra{#1}{#2}\else\sketbra{#1}{#2}\fi}

\newcommand{\proj}[1]{\ketbra{#1}{#1}}

\newcommand{\tp}{\otimes}
\newcommand{\idd}{\openone}

\newcommand{\tx}[1]{\textnormal{#1}}

\newcommand{\sotimes}{\scalebox{0.7}{$\otimes\,$}}

\newcommand{\rA}{\mathrm{A}}

\newcommand{\rB}{\mathrm{B}}

\newcommand{\rC}{\mathrm{C}}
\newcommand{\rD}{\mathrm{D}}

\newcommand{\A}{\mathsf{A}}
\newcommand{\B}{\mathsf{B}}
\newcommand{\M}{\mathsf{M}}

\newcommand{\X}{\mathsf{X}}
\newcommand{\Y}{\mathsf{Y}}
\newcommand{\Z}{\mathsf{Z}}
\newcommand{\D}{\mathsf{D}}
\newcommand{\E}{\mathsf{E}}
\newcommand{\G}{\mathsf{G}}
\newcommand{\Op}{\mathsf{O}}
\newcommand{\Pp}{\mathsf{P}}
\newcommand{\Ss}{\mathsf{S}}
\newcommand{\T}{\mathsf{T}}

\newcommand{\vz}{\mathbf{z}}
\newcommand{\vx}{\mathbf{x}}
\newcommand{\va}{\mathbf{a}}
\newcommand{\vb}{\mathbf{b}}
\newcommand{\vc}{\mathbf{c}}

\newcommand{\vw}{\mathbf{w}}
\newcommand{\vd}{\mathbf{d}}
\newcommand{\vy}{\mathbf{y}}

\newtheorem{theorem}{Theorem}
\newtheorem{lemma}[theorem]{Lemma}

\theoremstyle{definition}
\newtheorem{definition}{Definition}[section]

\begin{document}
\title{Self-testing of Pauli observables for device-independent entanglement certification}

\author{Joseph Bowles}
\affiliation{ICFO-Institut de Ciencies Fotoniques, The Barcelona Institute of Science and Technology, 08860 Castelldefels (Barcelona), Spain}
\author{Ivan \v{S}upi{\'c}}
\affiliation{ICFO-Institut de Ciencies Fotoniques, The Barcelona Institute of Science and Technology, 08860 Castelldefels (Barcelona), Spain}
\author{Daniel Cavalcanti}
\affiliation{ICFO-Institut de Ciencies Fotoniques, The Barcelona Institute of Science and Technology, 08860 Castelldefels (Barcelona), Spain}
\author{Antonio Ac{\'i}n}
\affiliation{ICFO-Institut de Ciencies Fotoniques, The Barcelona Institute of Science and Technology, 08860 Castelldefels (Barcelona), Spain}
\affiliation{ICREA - Instituci\'{o} Catalana de Recerca i Estudis Avancats, 08011 Barcelona, Spain}

\date{\today}

\begin{abstract}
We present self-testing protocols to certify the presence of tensor products of Pauli measurements on maximally entangled states of local dimension $2^n$ for $n\in\mathbb{N}$. The provides self-tests of sets of informationally complete measurements in arbitrarily high dimension. We then show that this can be used for the device-independent certification of the entanglement of all bipartite entangled states by exploiting a connection to measurement device-independent entanglement witnesses and quantum networks. This work extends a more compact parallel work on the same subject \cite{short} and provides all the required technical proofs.\end{abstract}

\maketitle

\section*{Introduction}
Unjustified or mistaken assumptions about the physics of a quantum information protocol can result in errors that jeopardise the protocol's validity \cite{rosset2012,makarov}. The \emph{device-independent} approach attempts to overcome this problem by keeping assumptions to a minimum; devices in the protocol are treated as black boxes, and the only information available is their input/output statistics. Interestingly, due to the existence of quantum nonlocality \cite{Bell,review}, protocols can still be made to function in this scenario and many quantum information tasks now have device-independent formulations, including protocols for quantum random number certification \cite{DIR1,DIR2,DIR3}, quantum key distribution \cite{DIQKD1,DIQKD2,DIQKD3} and the characterisation of quantum properties \cite{DIprot,DimWit}. 

A common device-independent task is that of \emph{entanglement certification}. Here, one aims to certify the presence of entanglement in a quantum state from the correlations between local measurement outcomes, and is typically achieved via the violation of a Bell inequality. The central limitation here is that there exist entangled mixed states that admit a so-called local hidden variable model \cite{Werner,Barrett,LHV} and thus do not violate any Bell inequality. Device-independent entanglement certification of such states is therefore impossible via the standard approach. A partial solution to this problem recently came in the form of measurement device-independent entanglement witnesses (MDIEWs) \cite{Buscemi,MDIEW1,MDIEW2}. Here, one can achieve entanglement certification of all entangled states by replacing the classical inputs in a Bell test by a set of trusted quantum input states. This approach, however, is only partially device-independent since it requires perfect knowledge of the input states. 

A closely related task to entanglement certification is that of \emph{self-testing} \cite{MY}. In a self-testing protocol, one aims to certify, or self-test, the presence of a target entangled state and/or target set of measurements via the observation of nonlocal correlations. Essentially, this requires finding a Bell inequality whose maximum violation is achieved uniquely by the target state and measurements of interest. A significant literature on self-testing exists \cite{YN,Bamps,SCAA,McKYS,McK,Wu,McK2}, and it is known for example that all bipartite pure entangled states can be self-tested \cite{CGS}. The self-testing of quantum measurements is however much less explored, although some results are known \cite{Jed,McKM}. 

In this work we combine results in the field of self-testing with techniques from MDIEWs to construct device-independent protocols that are capable of certifying the entanglement of all bipartite entangled states. To do this, we move to a scenario involving a network of quantum states that allows us to overcome the limitations of the standard approach. Intuitively, our protocols can be understood as a device-independent extension of MDIEWs, in which the input quantum states are certified device-independently via a self-testing protocol. The technical preliminaries to this result include new results concerning the parallel self-testing of Pauli observables and may be of independent interest. In particular, we  prove self-testing of tensor products of Pauli observables on maximally entangled states of local dimension $2^n$, $n\in\mathbb{N}$, treating a well known problem that arises when dealing with complex-valued measurements. We note that an analogous result to this was independently proven in \cite{leash} in the context of delegated quantum computation. 

The paper is organised as follows. The first two sections focus on the technical ground work in self-testing that are needed for our entanglement certification protocols. In section \ref{stsection} we introduce self-testing and revisit the problem that arises with complex-valued measurements. In section \ref{st2qubit}, we focus on the simplest case of two qubits and prove self-testing of the three Pauli observables, before tackling the more involved case of general dimension in sections \ref{parallelsection} and \ref{bsmsection} and discussing noise-robust versions of these results in section \ref{stnoise}. We then move to our protocols for entanglement certification, outlining our network scenario in section \ref{scenariosection}, presenting our entanglement certification protocols in section \ref{protocolsectionintro} - \ref{dinoise} and finally discussing our results. 

\section{Self-testing}\label{stsection}
Suppose two parties, Charlie and Alice\footnote{We avoid the usual convention of Alice and Bob for readability with later sections of this paper where our choice will become more natural.}, share the quantum state $\ket{\psi}$ and perform local measurements labelled by $z$ and $x$, obtaining outcomes $c$ and $a$. From the Born rule, the observed probabilities take the form 
\begin{align}\label{qprobs}
p(ca|zx)=\Tr\left[\proj{\psi}\M_{c|z}\otimes\M_{a|x}\right],
\end{align}
where $\M_{c|z}$, $\M_{a|x}$ denote the local measurement operators, and where we have purified states and measurements so that our state is a pure state and our measurements projective. In principle, many different combinations of states and measurements could give rise to the same correlations $p(ca|zx)$. To self-test a target quantum state $\ket{\psi'}$, one must find correlations which are produced uniquely by  $\ket{\psi'}$ up to a certain equivalence class, hence certifying the state $\ket{\psi'}$ (up to equivalence) from knowledge of the correlations alone. In the first works on self-testing, this equivalence class is captured by the notion of a local isometry, which takes into account the possibility of unobservable local unitary operations applied to the state and measurements, possible embedding in a Hilbert space of larger dimension and/or the existence of additional degrees of freedom. Note that via the Schmidt decomposition, the freedom of local unitary operations implies that one may assume that the target state $\ket{\psi'}$ can be expressed with real numbers only without loss of generality. The precise definition of self-testing of quantum states is then as follows. 

\begin{definition}\label{defststates}
We say that the correlations $p^*(ca\vert zx)$ self-test the state $\ket{\psi'}\in{\mathcal{H}_{\rC'}\otimes\mathcal{H}_{\rA'}}$ if for all states and all measurement operators satisfying \eqref{qprobs} for $p(ca|zx)=p^*(ca|zx)$ there exist Hilbert spaces $\mathcal{H}_{\rC}$, $\mathcal{H}_{\rA}$ such that $\ket{\psi}\in\mathcal{H}_{\rC}\otimes\mathcal{H}_{\rA}$, a local auxiliary state $\ket{00}\in\mathcal{H}_{\rC'}\otimes\mathcal{H}_{\rA'}$ and a local unitary operator $U$ such that
\begin{align}\label{selftest}
U \left[\ket{\psi}\otimes \ket{00}\right]=\ket{\xi}\otimes\ket{\psi'},
\end{align}
where $\ket{\xi}\in\mathcal{H}_{\rC}\otimes\mathcal{H}_{\rA}$ (usually called a junk state) is any state representing possible additional degrees of freedom.
\end{definition}
Intuitively, self-testing means proving the existence of local channels (given by the local unitaries and local auxiliary states) which extract the target state $\ket{\psi'}$ from the physical state $\ket{\psi}$ into the $\mathcal{H}_{\rC'}\otimes\mathcal{H}_{\rA'}$ space.

One may further be interested in certifying that the measurement operators are equivalent to some target measurements $\{\M'_{c|z}\}$, $\{\M'_{a|x}\}$ acting on $\ket{\psi'}$. To begin with, let us assume that the target measurements can be expressed using real numbers alone, i.e. $(\M'_{c|z})^*=\M'_{c|z}$ for all $c,z$ and $(\M'_{a|x})^*=\M'_{a|x}$ for all $a,x$. We then have the following definition.
\begin{definition}\label{defstmeas}
We say that the correlations $p^*(ca|zx)$ self-test the state $\ket{\psi'}$ and real-valued measurements $\{\M'_{c|z}\}$, $\{\M'_{a|x}\}$ if $p^*(ca|zx)$ self-tests the state $\ket{\psi'}$ according to definition\ \ref{defststates} and furthermore 
\begin{align}\label{selftest2}
U\left[\M_{c|z}\otimes\M_{a|x}\,\ket{\psi}\otimes \ket{00}\right]=\ket{\xi}\otimes(\M'_{c|z}\otimes\M'_{a|x}\ket{\psi'}) \nonumber
\end{align}
for each $c,a,z,x$. 
\end{definition}
In other words, applying the measurements $\M_{c|z}$, $\M_{a|x}$ to the state $\ket{\psi}$ is equivalent to applying $\M'_{c|z}$, $\M'_{a|x}$  to $\ket{\psi'}$ under the action of the local unitaries.

For measurements that cannot be expressed using real numbers alone an additional complication arises, as noted in the early works on self-testing \cite{McKThesis} (see also \cite{McKM} and \cite{Jed}). This is due to the fact that quantum correlations are invariant under transposition (or equivalently, complex conjugation) of the state and measurement operators:
\begin{equation}\label{transcor}
\Tr[\proj{\psi'}\M'_{c|z}\otimes\M'_{a|x}]=\Tr[\proj{\psi'}\M'^T_{c|z}\otimes\M'^T_{a|x}] 
\end{equation} 
(where $\M^T$ denotes the transposition operation and we assume the state $\ket{\psi'}$ to be real as above). Note that the transposition operation maps valid measurement operators to valid measurement operators, however is not unitary. This means that the measurements $\{\M'_{c|z}\}$, $\{\M'_{a|x}\}$ cannot be self-tested using definition \ref{defstmeas}. That is, there always exists an alternative realisation using the transposed measurements which cannot be brought to the target measurements using local isometries alone. For such measurements, the most we can hope to certify is that the measurement operators correspond to the target set up to the additional freedom of local transpositions on both subsystems. To deal with this possibility and following the method of \cite{McKM}, we introduce additional local Hilbert spaces $\mathcal{H}_{\rC''}$ and $\mathcal{H}_{\rA''}$ which act as a control space for possible transposition of the measurement operators. Our precise definition of self-testing is as follows. 
\begin{definition}\label{defstcomplex}
We say that the correlations $p^*(ca\vert zx)$ self-test the state $\ket{\psi'}\in{\mathcal{H}_{\rC'}\otimes\mathcal{H}_{\rA'}}$ and (complex-valued) measurements $\{\M'_{c|z}\}$, $\{\M'_{a|x}\}$ if for all states and all measurement operators satisfying \eqref{qprobs} for $p(ca|zx)=p^*(ca|zx)$ there exist Hilbert spaces $\mathcal{H}_{\rC}$, $\mathcal{H}_{\rA}$ such that $\ket{\psi}\in\mathcal{H}_{\rC}\otimes\mathcal{H}_{\rA}$, a local auxiliary state $\ket{00}\in[\mathcal{H}_{\rC''}\otimes\mathcal{H}_{\rC'}]\otimes[\mathcal{H}_{\rA''}\otimes\mathcal{H}_{\rA'}]$ and a local unitary operator $U$ such that
\begin{multline}\label{selftestfull} 
U\left[\M_{c|z}\otimes\M_{a|x}\,\ket{\psi}\otimes \ket{00}\right]=\\\tilde{\M}_{c|z}\otimes\tilde{\M}_{a|c}\left[\ket{\xi_0}\otimes\ket{00}+\ket{\xi_1}\otimes\ket{11}\right]\otimes\ket{\psi'},
\end{multline}
where $\ket{\xi_j}\in\mathcal{H}_{\rC}\otimes\mathcal{H}_{\rA}$ are some unknown sub-normalised junk states such that $\langle \xi_0 \vert  \xi_0 \rangle+\langle \xi_1 \vert  \xi_1 \rangle=1$ and the $\tilde{\M}$ operators are related to the target measurements by
\begin{align}
\tilde{\M}_{c|z}&=\openone^{\rC}\otimes\left[\M_0\otimes\M'_{c|z}+\M_1\otimes(\M'_{c|z})^T\right]; \\
\tilde{\M}_{a|x}&=\openone^{\rA}\otimes\left[\M_0\otimes\M'_{a|x}+\M_1\otimes(\M'_{a|x})^T\right],
\end{align}
with $\M_0+\M_1=\openone^{\rC''}$ and $\langle0\vert\M_0\vert0\rangle=\langle1\vert\M_1\vert1\rangle=1$.
\end{definition}
The above measurements can be understood as `controlled transposition' measurements: one first measures the double primed auxiliary spaces with the measurement $\{\M_0,\M_1\}$; conditioned on this outcome, one then measures the target measurement or its transposition on the target state $\ket{\psi'}$. Due to the form of the measurement operators and state $\ket{\xi_0}\otimes\ket{00}+\ket{\xi_1}\otimes\ket{11}$, one sees that this transposition is correlated between Charlie and Alice, as implied from \eqref{transcor}. The probability that this transposition is applied depends on the norm of the vectors $\ket{\xi_j}$, however is generally unknown since the self-testing data does not allow one to infer the form of these states. Note that one may only wish to self-test a set of measurements for one of the parties, say Charlie (as will be the case for us); here one would simply replace the measurement operators for Alice by the identity operator in the above.

The central task in self-testing is thus to construct the local unitary $U$ in order to prove statements following the above definitions. In order to do this, one typically considers linear combinations of the probabilities $p(ca|zx)$ (corresponding to some Bell inequality) of the form 
\begin{align}\label{bellin}
\mathcal{I}\left[p(ca|zx)\right]=\sum_{c,a,z,x} \beta^{zx}_{ca}\,p(ca|zx), 
\end{align}
for which the maximal value in quantum theory $\mathcal{I}=\mathcal{I}_{\tx{max}}$ occurs using the target state and measurements. The observation $\mathcal{I}=\mathcal{I}_{\tx{max}}$ then implies relations between the state and measurements performed in the experiment via \eqref{bellin}, and one can prove the existence of the local unitary from the measurement operators themselves. A large number of self-testing results are known. For example, if \eqref{bellin} corresponds to the CHSH Bell inequality, maximum violation implies that one can self-test the presence of a maximally entangled state of dimension two $\ket{\Phi^{\tx{+}}}=\frac{1}{\sqrt{2}}[\ket{00}+\ket{11}]$, and measurements of $\sigma_{\tx{x}}$, $\sigma_{\tx{z}}$ for Charlie and $[\sigma_{\tx{x}}\pm\sigma_{\tx{z}}]/\sqrt{2}$ for Alice \cite{MY,McKYS,Kaniewski}. More generally, one can self-test any pure bipartite entangled two-qubit state $\ket{\psi} = \cos{\theta}\ket{00} + \sin{\theta}\ket{11}$ when \eqref{bellin} corresponds to the tilted CHSH Bell inequality \cite{Bamps}. Self-testing of higher dimensional bipartite pure states is given in \cite{CGS,McK,McK2,Wu}. Furthermore, a large class of multipartite states can be self-tested by exploiting the methods applied to self-testing of bipartite states \cite{SCAA}. 

The majority of self-tests mentioned above are useful for the certification of measurements. However, most of these results apply to the self-testing of real-valued measurements  due to the added complication definition \ref{defstcomplex}. The simplest set of measurements which cannot be expressed using real numbers alone is given by the three Pauli observables $\sigma_{\tx{z}},\sigma_{\tx{x}},\sigma_{\tx{y}}$. In Section \ref{st2qubit} we prove self-testing statements for these measurements, inspired by the approach of \cite{McKM} where similar results were obtained. We then extend this to a parallel self-test in Sections \ref{parallelsection} and \ref{bsmsection} in order to prove self-testing statements for $n$-fold tensor products of the Pauli measurements, which form an informationally complete set in dimension $2^n$.

\subsection{Self-testing of Pauli measurements}\label{st2qubit}
\begin{figure*}
\centering
\includegraphics[scale=1]{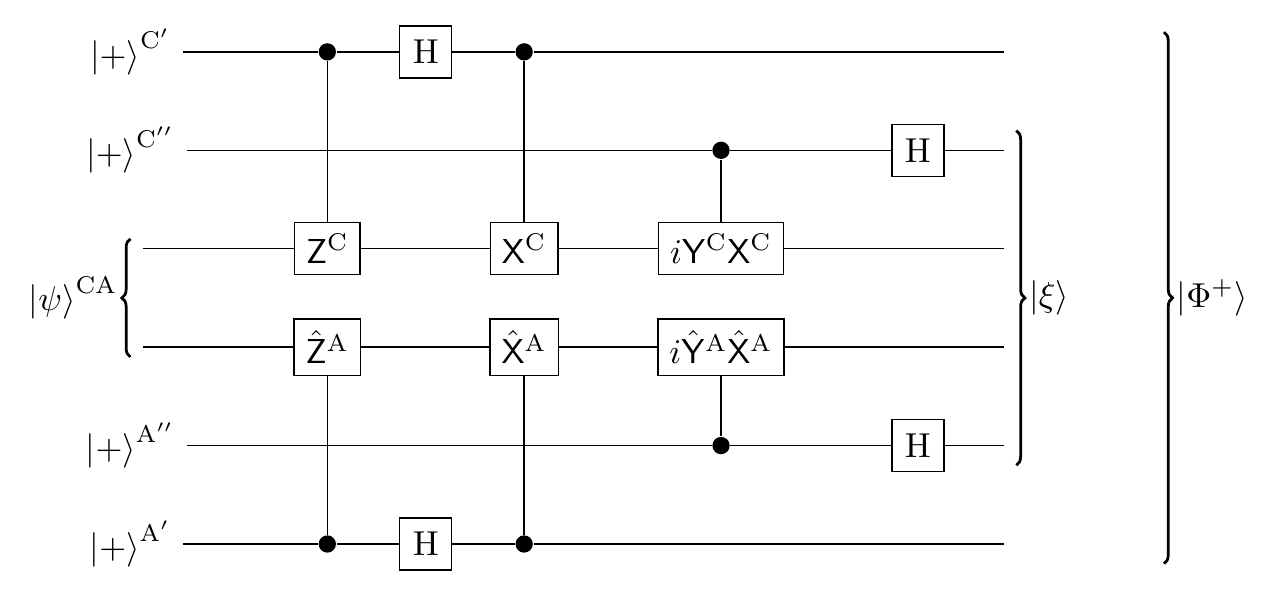}
\caption{Self-testing circuit used for the proof of Lemma \ref{lemma1}. The unitaries $\hat{\Z}^{\rA}$, $\hat{\X}^{\rA}$, $\hat{\Y}^{\rA}$ can be found in appendix \ref{lemma1proof}.  }
\label{fig:STcircuit}
\end{figure*}

We begin by proving a self-testing statement for the maximally entangled state of two qubits $\ket{\Phi^{\tx{+}}}=\frac{1}{\sqrt{2}}[\ket{00}+\ket{11}]$ and the three Pauli observables for Charlie. Since there does not exist a two-qubit basis in which these observables can be written using real numbers only, our self-testing statement will be of the form of definition \ref{defstcomplex}. We note that this is not the first proof of such a result; similar results have been obtained in previous works by generalising the Mayers-Yao self-test \cite{McKM}, by studying the properties of the `elegant' Bell inequality \cite{Toni,Cabello} and combinations of the CHSH Bell inequality \cite{Toni} and in a more general approach to the problem \cite{Jed} focused on commutation relations. 

Before proceeding we first clarify some notation. Superscript of an operator denotes the Hilbert space on/in which the operator acts/lives, e.g.\ $\X^\rC$ denotes a linear operator on the space $\mathcal{H}_{\rC}$ and $\ket{\psi}^{\rC\rA}\in\mathcal{H}_{\rC}\otimes\mathcal{H}_{\rA}$. Unless explicitly written, we omit tensor products acting on the remaining Hilbert space, e.g.\ $\X^\rC\ket{\psi}^{\rC\rA}$ should be understood as $\X^\rC\otimes\idd^{\rA}\ket{\psi}^{\rC\rA}$. This convention then follows for the product of operators, e.g.\ $\X^\rC\E^\rA\ket{\psi}^{\rC\rA}$ should be understood as $\X^\rC\otimes\E^\rA\ket{\psi}^{\rC\rA}$. 

The scenario we consider for the self-testing is as follows. Charlie and Alice share a bipartite quantum state $\ket{\psi}\in\mathcal{H}_{\rC}\otimes\mathcal{H}_{\rA}$. Charlie has a choice of three measurements $z=1,2,3$, with outcomes $c=\pm1$ denoted by the observables $\X^\rC, \Y^\rC$ and $\Z^\rC$. Alice has a choice of six $\pm1$ valued measurements $x=1,\cdots,6$, $a=\pm1$, denoted by the observables $\D_{\tx{z,x}}^{\rA}, \E_{\tx{z,x}}^{\rA},\D_{\tx{x,y}}^{\rA}, \E_{\tx{x,y}}^{\rA},\D_{\tx{z,y}}^{\rA}, \E_{\tx{z,y}}^{\rA}$. Note that each of these observables is Hermitian and unitary. We then consider the following Bell operator (introduced in \cite{Toni}), which we call the triple CHSH Bell operator
\begin{multline}\label{bellop}
\mathcal{B}=\Z^{\rC} (\D_{\tx{z,x}}^{\rA} + \E_{\tx{z,x}}^{\rA}) + \X^\rC (\D_{\tx{z,x}}^{\rA} - \E_{\tx{z,x}}^{\rA}) \\
+ \Z^\rC (\D_{\tx{z,y}}^{\rA} + \E_{\tx{z,y}}^{\rA}) - \Y^\rC (\D_{\tx{z,y}}^{\rA} - \E_{\tx{z,y}}^{\rA}) \\
+ \X^\rC (\D_{\tx{x,y}}^{\rA} + \E_{\tx{x,y}}^{\rA}) - \Y^\rC (\D_{\tx{x,y}}^{\rA} - \E_{\tx{x,y}}^{\rA}).
\end{multline}
This Bell operator consists of a sum of three CHSH Bell operators; each line itself is a CHSH Bell operator and each $\X$, $\Y$ and $\Z$ observable appears in two of the lines. The correlations that we use for self-testing correspond to those which maximise $\bra{\psi}\mathcal{B}\ket{\psi}$, which has maximum value $6\sqrt{2}$ (since each CHSH operator is upper bounded by $2\sqrt{2}$). This can be achieved by taking the following states and observables 
\begin{align}\label{qubitstrategy}
\ket{\psi}=\ket{\Phi^{+}}=\frac{1}{\sqrt{2}}\left[\ket{00}+\ket{11}\right], \nonumber \\
\Z^{C}=\sigma_{\tx{z}}\,, \quad \X^{C}=\sigma_{\tx{x}}\,, \quad \Y^{C}=\sigma_{\tx{y}}, \nonumber \\
\D_{i,j}^{\rA}=\frac{\sigma_{i}+\sigma_{j}}{\sqrt{2}}\,, \quad \E_{i,j}^{\rA}=\frac{\sigma_{i}-\sigma_{j}}{\sqrt{2}},
\end{align}
for $(i,j)=(\tx{z,x}),(\tx{z,y}),(\tx{x,y})$. The basic intuition of the self-testing is that since maximal violation of a single CHSH inequality requires anti-commuting qubit observables on a maximally entangled state \cite{CHSH}, the maximum value of \eqref{bellop} should imply three mutually anti-commuting observables on the maximally entangled state, given by the three Pauli observables (or their transpositions). Indeed, we will see that this is the case. 

One way to achieve this is to build a sum-of-squares (SOS) decomposition of the shifted Bell operator $6\sqrt{2}\openone-\mathcal{B}$ of the form
\begin{align}\label{SOS}
6\sqrt{2}\openone-\mathcal{B}=\sum_{\lambda}P_{\lambda}^{\dagger}P_{\lambda}.
\end{align}
Such a decomposition is given by
\begin{align}\label{chshsos}
&2\left(6\sqrt{2}\openone-\mathcal{B}\right)=\\
&\left[\Z^\rC - \scalebox{1}{$\frac{1}{\sqrt{2}}$}(\D_{\tx{z,x}}^{\rA} + \E_{\tx{z,x}}^{\rA})\right]^2 + \left[ \X^\rC - \scalebox{1}{$\frac{1}{\sqrt{2}}$}(\D_{\tx{z,x}}^{\rA} - \E_{\tx{z,x}}^{\rA})\right]^2 \nonumber\\ + &\left[ \Z^\rC -\scalebox{1}{$\frac{1}{\sqrt{2}}$}(\D_{\tx{z,y}}^{\rA} + \E_{\tx{z,y}}^{\rA})\right]^2  + \left[\Y^\rC + \scalebox{1}{$\frac{1}{\sqrt{2}}$}(\D_{\tx{z,y}}^{\rA} - \E_{\tx{z,y}}^{\rA})\right]^2   \nonumber \\ +&\left[\X^\rC - \scalebox{1}{$\frac{1}{\sqrt{2}}$}(\D_{\tx{x,y}}^{\rA} + \E_{\tx{x,y}}^{\rA})\right]^2 + \left[\Y^\rC + \scalebox{1}{$\frac{1}{\sqrt{2}}$}(\D_{\tx{x,y}}^{\rA} - \E_{\tx{x,y}}^{\rA})\right]^2 \nonumber.
\end{align}\normalsize
Here, the $P_{\lambda}$'s are Hermitian and so $P_{\lambda}^{\dagger}P_{\lambda}=P_{\lambda}^2$. At maximal value one has $\bra{\psi}\mathcal{B}\ket{\psi}=6\sqrt{2}$ and so
\begin{align}
\sum_{\lambda}\bra{\psi}P_{\lambda}^\dagger P_{\lambda}\ket{\psi}=0.
\end{align}
Since each term in the above is greater or equal to zero we have $P_{\lambda}\ket{\psi}=0$ for all $\lambda$. Applying this to the SOS decomposition \eqref{chshsos} gives
\begin{align}\label{xyzcons1}
\Z^\rC\ket{\psi} &= \scalebox{1}{$\frac{1}{\sqrt{2}}$}[\D_{\tx{z,x}}^{\rA} + \E_{\tx{z,x}}^{\rA}]\ket{\psi} = \scalebox{1}{$\frac{1}{\sqrt{2}}$}[\D_{\tx{z,y}}^{\rA} + \E_{\tx{z,y}}^{\rA}]\ket{\psi},\\ 
\X^\rC\ket{\psi} &= \scalebox{1}{$\frac{1}{\sqrt{2}}$}[\D_{\tx{z,x}}^{\rA} - \E_{\tx{z,x}}^{\rA}]\ket{\psi} = \scalebox{1}{$\frac{1}{\sqrt{2}}$}[\D_{\tx{x,y}}^{\rA} + \E_{\tx{x,y}}^{\rA}]\ket{\psi},\\
\Y^\rC\ket{\psi} &= \scalebox{1}{$\frac{1}{\sqrt{2}}$}[\E_{\tx{z,y}}^{\rA} - \D_{\tx{z,y}}^{\rA}]\ket{\psi} = \scalebox{1}{$\frac{1}{\sqrt{2}}$}[\E_{\tx{x,y}}^{\rA} - \D_{\tx{x,y}}^{\rA}]\ket{\psi}. \label{xyzcons3}
\end{align}
Since for any two unitary observables $\G_1$ and $\G_2$, the composite observables $\frac{\G_1+\G_2}{\sqrt{2}}$ and $\frac{\G_1-\G_2}{\sqrt{2}}$ anti-commute by construction, from the above three equations it follows that on the support of state $\ket{\psi}$ observables $\Z^\rC,\X^\rC$ and $\Y^\rC$ mutually anti-commute:
\be\label{ac}
\{\Z^\rC,\X^\rC\}\ket{\psi} = \{\Z^\rC,\Y^\rC\}\ket{\psi} = \{\X^\rC,\Y^\rC\}\ket{\psi} = 0. 
\ee
The conditions \eqref{xyzcons1} - \eqref{xyzcons3} and \eqref{ac} allow us to construct a local unitary which will give us our desired self-testing. This unitary can be understood via the circuit of Fig.\ \ref{fig:STcircuit}, and is based on the swap gate introduced in \cite{McKYS} and is the same as the circuit found in \cite{McKM}. The unitaries $\hat{\Z}^{\rA}$, $\hat{\X}^{\rA}$, $\hat{\Y}^{\rA}$ are regularized versions of the operators 
\small\begin{align}
{\Z}^{\rA}=\frac{\D_{\tx{z,x}}^{\rA} + \E_{\tx{z,x}}^{\rA}}{\sqrt{2}}, \quad {\X}^{\rA}=\frac{\D_{\tx{z,x}}^{\rA} - \E_{\tx{z,x}}^{\rA}}{\sqrt{2}}, \quad {\Y}^{\rA}=\frac{\E_{\tx{z,y}}^{\rA} - \D_{\tx{z,y}}^{\rA}}{\sqrt{2}} \nonumber.
\end{align}\normalsize
For example, $\hat{\Z}^{\rA}$ is obtained by setting all zero eigenvalues of ${\Z}^{\rA}$ to one and then defining $\hat{\Z}^{\rA}={\Z}^{\rA}\vert{\Z}^{\rA}\vert^{-1}$. Using standard techniques (see appendix \ref{lemma1proof}), these can be shown to act in the same way as the non-regularised versions. With this we are ready to present the first of our self-testing lemmas. 
\begin{tcolorbox}
\begin{lemma}\label{lemma1}
Let the state $\ket{\psi}\in\mathcal{H}_\rC\otimes\mathcal{H}_{\rA}$ and $\;\pm1\;$ outcome observables $\X^{\rC}$, $\Y^{\rC}$, $\Z^{\rC}$, $\D_{\tx{z,x}}^{\rA}$, $\E_{\tx{z,x}}^{\rA}$, $\D_{\tx{x,y}}^{\rA}$, $\E_{\tx{x,y}}^{\rA}$, $\D_{\tx{z,y}}^{\rA}$, $\E_{\tx{z,y}}^{\rA}$ satisfy
\begin{align}\label{tripleCHSH} 
&\bra{\psi} \mathcal{B}\ket{\psi} = 6\sqrt{2}.
\end{align}
Then there exist local auxiliary states $\ket{00}\in [\mathcal{H}_{\rC''}\otimes\mathcal{H}_{\rC'}] \otimes [\mathcal{H}_{\rA''}\otimes\mathcal{H}_{\rA'}]$ and a local unitary $U$ such that:
\begin{align} \label{ststate}
U[\ket{\psi}\otimes\ket{00}] &= \ket{\xi}\tp \ket{\Phi^{\tx{+}}}^{\rC'\rA'},\\ \label{stx}
U[\X^\rC\ket{\psi}\otimes\ket{00}] &= \ket{\xi} \tp \sigma_{\tx{x}}^{\rC'}\ket{\Phi^{\tx{+}}}^{\rC'\rA'},\\ \label{stz}
U[\Z^\rC\ket{\psi}\otimes\ket{00}] &= \ket{\xi}\tp \sigma_{\tx{z}}^{\rC'}\ket{\Phi^{\tx{+}}}^{\rC'\rA'},\\ \label{sty}
U[\Y^\rC\ket{\psi}\otimes\ket{00}] &= \sigma_{\tx{z}}^{\rC''}\ket{\xi}\tp \sigma_{\tx{y}}^{\rC'}\ket{\Phi^{\tx{+}}}^{\rC'\rA'}, 
\end{align}
where $\ket{\xi}$ takes the form 
\begin{align}\label{junk}
\ket{\xi} =\ket{\xi_0}^{\rC\rA}\otimes\ket{00}^{\rC''\rA''} + \ket{\xi_1}^{\rC\rA}\otimes\ket{11}^{\rC''\rA''}. 
\end{align}
\end{lemma}
\end{tcolorbox}
Note that the complex observable $\sigma_{\tx{y}}$ has an additional $\sigma_{\tx{z}}$ measurement on the $\rC''$ space, as expected from definition \ref{defstcomplex}.  Hence, the measurement $\Y$ can be understood as first measuring $\sigma_\tx{z}$ on the state $\ket{\xi}$, whose outcome decides whether $\pm\sigma_{\tx{y}}$ is performed on the state $\ket{\Phi^{\tx{+}}}$. The probability that the observables $\{\sigma_{\tx{x}},\sigma_{\tx{y}},\sigma_{\tx{z}}\}$ are used rather than the transposed measurements $\{\sigma_{\tx{x}},-\sigma_{\tx{y}},\sigma_{\tx{z}}\}$ is given by the probability to obtain $+1$ for the $\sigma_{\tx{z}}^{\rC''}$ measurement. As mentioned in Section \ref{stsection}, this probability remains unknown since one does not know the precise form of $\ket{\xi}$ from the self-testing correlations alone. The proof of Lemma 1 can be found in appendix \ref{lemma1proof}.

\subsection{Parallel self-testing of Pauli observables}\label{parallelsection}

The protocol described above can be extended to a parallel self-test. Here, our aim is to self-test the $n$-fold tensor product of the maximally entangled state $\ket{\Phi^{\tx{+}}}^{\otimes n}$ (which itself is a maximally entangled state of dimension $2^n$) and all combinations of $n$-fold tensor products of Pauli measurements for Charlie, i.e.\ $\sigma_{i_1}\otimes\sigma_{i_2}\otimes\cdots\otimes\sigma_{i_n}$ for $i_j=\tx{x},\tx{y},\tx{z}$. This is achieved by an $n$-fold maximal parallel violation of the Bell inequality used in Lemma \ref{lemma1}. As a basis we use the techniques of \cite{Col}, where parallel self-testing of $\sigma_{\tx{x}}$ and $\sigma_{\tx{z}}$ observables on the maximally entangled state was proven. Besides \cite{Col}, parallel self-testing of $n$-fold tensor products of maximally entangled pairs of qubits has been presented in \cite{McK} and \cite{CN}, and in \cite{Wu} for $n=2$. This section can thus be seen as an extension of these results to all three Pauli observables. Although we use the term `self-testing' here, we will see that simply performing the protocol of Lemma \ref{lemma1} in parallel does not lead to a self-test according to definition \ref{defstcomplex}. In the following subsection we  correct this by adding additional Bell state measurements between local subsystems. 

The scenario we consider is as follows. Charlie and Alice share the state $\ket{\psi}\in\mathcal{H}_\rC\otimes\mathcal{H}_\rA$. Charlie has a choice of $3^n$ measurements collected into the vector $\mathbf{z}=(z_1,z_2,\cdots,z_n)$ with $z_i=1,2,3$, and each measurement has $2^n$ possible outcomes given by $\mathbf{c}=(c_1,c_2,\cdots,c_n)$ with $c_i=\pm1$. Similarly, Alice has a choice of $6^n$ measurements given  by the vector $\mathbf{x}=(x_1,x_2,\cdots,x_n)$ with $x_i=1,2,3,4,5,6$, each with $2^n$ possible outputs given by $\va=(a_1,a_2,\cdots,a_n)$ with $a_i=\pm1$. Fixing a value of $i$ we thus have three possible settings for Charlie and six for Alice, corresponding to the self-test of the previous section that we now perform in parallel. In order to achieve this we will define an analogous Bell operator to \eqref{bellop} for each value of $i$.  

To this end, we denote Charlie's and Alice's measurement projectors by $\Pi^{\rC}_{\mathbf{c}|\mathbf{z}}$ and $\Pi^{\rA}_{\mathbf{a}|\mathbf{x}}$ respectively. We then define the following unitary observables for Charlie
\begin{align}
\Op_{i|\vz}=\sum_{\vc|c_i=+1}\Pi_{\vc|\vz}^{\rC}-\sum_{\vc|c_i=-1}\Pi_{\vc|\vz}^{\rC}.
\end{align}
These operators can be understood as $\pm1$ valued observables that depend on the output $c_i$ only for a particular choice of input $\vz$, and are thus analogous to one of the three Pauli measurements (given by the value $z_i$) acting on the $i^{th}$ subspace of the maximally entangled state. Next we define the operators 
\begin{align}\label{zop}
\Z^{\rC}_i=\frac{1}{3^{n-1}}\sum_{\vz|z_i=1}\Op_{i|\vz}, \\
\X^{\rC}_i=\frac{1}{3^{n-1}}\sum_{\vz|z_i=2}\Op_{i|\vz},\label{xop}\\
\Y^{\rC}_i=\frac{1}{3^{n-1}}\sum_{\vz|z_i=3}\Op_{i|\vz} \label{yop},
\end{align}
that is, the average observables compatible with a particular choice of $z_i$.

Similarly for Alice we define the unitary observables
\begin{align}
\Pp_{i|\vx}=\sum_{\va|a_i=+1}\Pi_{\va|\vx}^{\rA}-\sum_{\va|a_i=-1}\Pi_{\va|\vx}^{\rA}
\end{align}
and the six operators
\begin{align}
\D_{\tx{zx},i}^{\rA}=\frac{1}{6^{n-1}}\sum_{\vx|x_i=1}\Pp_{i|\vx} , \quad 
\E_{\tx{zx},i}^{\rA}=\frac{1}{6^{n-1}}\sum_{\vx|x_i=2}\Pp_{i|\vx},\nonumber \\ 
\D_{\tx{zy},i}^{\rA}=\frac{1}{6^{n-1}}\sum_{\vx|x_i=3}\Pp_{i|\vx} , \quad 
\E_{\tx{zy},i}^{\rA}=\frac{1}{6^{n-1}}\sum_{\vx|x_i=4}\Pp_{i|\vx}, \nonumber \\
\D_{\tx{xy},i}^{\rA}=\frac{1}{6^{n-1}}\sum_{\vx|x_i=5}\Pp_{i|\vx} , \quad
\E_{\tx{xy},i}^{\rA}=\frac{1}{6^{n-1}}\sum_{\vx|x_i=6}\Pp_{i|\vx}. \label{exop}
\end{align}
We now consider Bell operators of the form 
\begin{multline}
\mathcal{B}_i=\Z^\rC_i(\D^\rA_{\tx{zx},i}+\E^\rA_{\tx{zx},i})+\X^\rC_i(\D^\rA_{\tx{zx},i}-\E^\rA_{\tx{zx},i}) + \\ + \Z^\rC_i(\D^\rA_{\tx{zy},i}+\E^\rA_{\tx{zy},i})-\Y^\rC_i(\D^\rA_{\tx{zy},i}-\E^\rA_{\tx{zy},i}) \\  +\X^\rC_i(\D^\rA_{\tx{xy},i}+\E^\rA_{\tx{xy},i})-\Y^\rC_i(\D^\rA_{\tx{xy},i}-\E^\rA_{\tx{xy},i}).
\end{multline}
This is simply the Bell inequality \eqref{bellop}, for the inputs $z_i$ and $x_i$ averaged over all compatible $\vz$ and $\vx$. One can thus obtain $\bra{\psi}\mathcal{B}_i\ket{\psi}=6\sqrt{2}$ for each $i$ by taking $n$ copies of the maximally entangled state of dimension two and adopting the previous measurement strategy \eqref{qubitstrategy} independently on each of the copies. From the observation of maximal violation for all $i$, a self-testing circuit (a parallel version of the circuit of Lemma 1) can be constructed, see Fig.\ \ref{circuitgen} in appendix \ref{lemma2proof}. We then have the following lemma.
\begin{tcolorbox}
\begin{lemma}\label{lemma2}  Let the state $\ket{\psi}\in\mathcal{H}_\rC\otimes\mathcal{H}_\rA$ and operators $\Z_i^\rC$, $\X_i^\rC$, $\Y_i^\rC$, $\D_{\tx{zx},i}^{\rA}$, $\E_{\tx{zx},i}^{\rA}$, $\D_{\tx{zy},i}^{\rA}$, $\E_{\tx{zy},i}^{\rA}$, $\D_{\tx{xy},i}^{\rA}$, $\E_{\tx{xy},i}^{\rA}$ defined above satisfy
\begin{equation}\label{optCorr}
\bra{\psi}\mathcal{B}_i\ket{\psi} = 6\sqrt{2},
\end{equation}
for every $i \in \{1,\dots n\}$. Then there exists a local unitary $U$, local registers $\ket{00}\in \otimes_{i=1}^n[\mathcal{H}_{\rC_i''}\otimes\mathcal{H}_{\rC_i'}]\otimes[\mathcal{H}_{\rA_i''}\otimes\mathcal{H}_{\rA_i'}]$ and a normalized state $\ket{\xi}$ such that
\ba\label{1} \nonumber
U\left[\ket{\psi}\otimes\ket{00}\right] &=& \ket{\xi} \otimes \left[ \tp_{i=1}^n {\ket{\Phi^{\tx{+}}}}^{\rC_i'\rA_i'}\right],\\ \nonumber
U\left[\Z_j^\rC\ket{\psi}\otimes\ket{00}\right]&=& \ket{\xi} \otimes\left[ {\sigma_{\tx{z}}}^{\rC_j'} \tp_{i=1}^n{\ket{\Phi^{\tx{+}}}}^{\rC_i'\rA_i'}\right],\\ \nonumber
U\left[\X_j^\rC\ket{\psi}\otimes\ket{00}\right]&=& \ket{\xi}\otimes\left[{\sigma_{\tx{x}}}^{\rC_j'}\tp_{i=1}^n{\ket{\Phi^{\tx{+}}}}^{\rC_i'\rA_i'}\right],\\ \nonumber
U\left[\Y_j^\rC\ket{\psi}\otimes\ket{00}\right]&=& {\sigma_{\tx{z}}}^{\rC_j''}\ket{\xi}\otimes\left[{\sigma_{\tx{y}}}^{\rC_j'}\tp_{i=1}^n{\ket{\Phi^{\tx{+}}}}^{\rC_i'\rA_i'}\right], \nonumber
\ea
for every $j \in \{1,2,\dots n\}$, where $\ket{\xi}$ takes the form
\begin{align}\label{junk2}
\ket{\xi} = \sum_{\bar{q}}\ket{\xi_{\bar{q}}}^{\rC\rA}\otimes\ket{\bar{q}\bar{q}}^{\rC''\rA''} 
\end{align}
and the sum is over all bit strings $\bar{q}=(0,1)^n$
\end{lemma}
\end{tcolorbox}
The proof of the above Lemma can be found in the appendix \ref{lemma2proof}. Note that since the self-tested measurements are extremal then the above statement must hold not only for the operators $\Z_j$, $\X_j$, $\Y_j$ but for each of the observables $\Op_{i|\vz}$ appearing in their definition, which implies that the input $z_i$ indeed corresponds to the desired Pauli measurement on the correct subspace. The measurement ${\sigma_{\tx{z}}}^{\rC_j''}$ on the state $\ket{\xi}$ again plays the role of deciding whether the measurement $\sigma_{\tx{y}}^{\rC_j'}$ or $-\sigma_{\tx{y}}^{\rC_j'}$ is performed on the maximally entangled state. However, note that due to the form of $\ket{\xi}$, this is not guaranteed to be correlated with the other measurements of $\sigma_{y}$ on different local subspaces. As a result, one cannot equate this freedom to a local transposition on \emph{all} of Charlie's subsystems, as needed from definition \ref{defstcomplex}. In the following section we show how to overcome this problem by introducing additional measurement for Alice.

\begin{figure}
\includegraphics[scale=1.3]{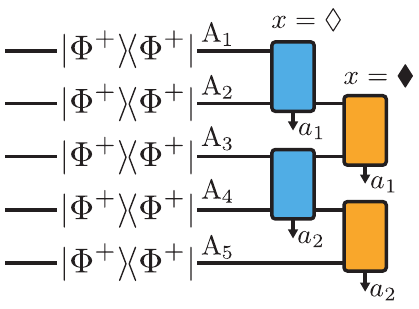}
\caption{\label{bsmfig} Graphical representation of the additional measurements performed by Alice for $x=\lozenge$ and $x=\blacklozenge$. Boxes between subspaces represent Bell state measurements.}
\end{figure}
\subsection{Aligning reference frames} \label{bsmsection}
As mentioned, Lemma \ref{lemma2} suffers from one drawback, namely that the $\tx{y}$ direction for each of Charlie's local subsystems need not be aligned. For example, if we take the case $n=2$, Lemma \ref{lemma2} gives four possibilities for Charlie's effective measurements on the maximally entangled state given by $\{\sigma_{\tx{x}},\pm\sigma_{\tx{y}},\sigma_{\tx{z}}\}\otimes\{\sigma_{\tx{x}},\pm\sigma_{\tx{y}},\sigma_{\tx{z}}\}$. The probability that each of these strategies is used is unknown and could, for example, be $\frac{1}{4}$ for each. In this case, when the first subsystem measures $\sigma_{\tx{y}}$, the second subsystem has equal probability to measure either $\sigma_{\tx{y}}$ or $-\sigma_{\tx{y}}$. This lack of alignment is an artefact from performing the protocol of Lemma \ref{lemma1} in parallel without trying to introduce any dependencies between the $n$ individual self-tests. In the following we show that one can further restrict the the state $\ket{\xi}$ to be of the form
\begin{align}\label{newxi}
\ket{\xi}=\ket{\xi_0}\otimes\ket{00\cdots 0}^{\rC''\rA''}+\ket{\xi_1}\otimes\ket{11\cdots 1}^{\rC''\rA''}
\end{align}
by introducing additional Bell state measurements between subsystems of Alice. Since $\ket{\xi}$ now has only two terms, the flipping of the $\sigma_{\tx{y}}$ measurements is always correlated; either none of the  measurements are flipped (each subsystem measures $\sigma_{\tx{y}}$) or all the measurements are flipped (each subsystem measures $-\sigma_{\tx{y}}$). We note that an analogous result was independently obtained in \cite{leash} (see Lemma 8 therein) using a similar approach.

To illustrate the basic idea let us again consider the case $n=2$, and assume we adopt the ideal measurement strategy (i.e. the strategy \eqref{qubitstrategy} in parallel). We now add an additional Bell state measurement for Alice which she performs on her two halves of the maximally entangled states. If Alice receives the outcome corresponding to the projector $\proj{\Phi^{\tx{+}}}$, via entanglement swapping Charlie will hold the state $\ket{\Phi^{\tx{+}}}$ in his local subsystem (for the other outcomes he will hold a different Bell state). This state has correlations $\bra{\Phi^{\tx{+}}}\sigma_{\tx{x}}\otimes\sigma_{\tx{x}}\ket{\Phi^{\tx{+}}}=+1$, $\bra{\Phi^{\tx{+}}}\sigma_{\tx{y}}\otimes\sigma_{\tx{y}}\ket{\Phi^{\tx{+}}}=-1$, $\bra{\Phi^{\tx{+}}}\sigma_{\tx{z}}\otimes\sigma_{\tx{z}}\ket{\Phi^{\tx{+}}}=+1$. Hence, in order to reproduce these correlations, the direction of Charlie's two measurements of $\sigma_{\tx{y}}$ need to be correlated as otherwise we would not have perfect anti-correlation for the measurement $\sigma_{\tx{y}}\otimes\sigma_{\tx{y}}$. In the following we formalise this intuition to strengthen Lemma \ref{lemma2} so that $\ket{\xi}$ is of the form \eqref{newxi}.

The precise scenario we consider is the following. In addition to the $6^n$ measurements of Lemma \ref{lemma2} given by the vector $\vx$, Alice has two extra measurements denoted by $x=\lozenge$ and $x=\blacklozenge$. These measurements have respectively $4^{m}$ and $4^{m'}$ outcomes, where $m=\lfloor\frac{n}{2}\rfloor$ and $m'=\lfloor\frac{n-1}{2}\rfloor$, which are grouped into the vectors $\va=(a_1,a_2,\cdots, a_m)$ and $\va=(a_1,a_2,\cdots, a_{m'})$ with $a_i=0,1,2,3$. We denote by $\Pi_{\va,\lozenge}$ and $\Pi_{\va,\blacklozenge}$ the projectors corresponding to the outcomes of these measurements and define the projectors for $l=1,\cdots,n$
\begin{align}
\Ss_{l,a^*}=\sum_{\va:a_l=a^*}\Pi_{\va|\lozenge} , \quad \T_{l,a^*}=\sum_{\va:a_l=a^*}\Pi_{\va|\blacklozenge},
\end{align}
that is, the projectors onto the the subspace corresponding to $a_l=a^*$ for the two measurements.

\def\arraystretch{1.5}
\begin{table}\label{bsmtab1}
\begin{tabular}{ |c | c | c | c | c |} 
\hline
 & $\quad \idd \quad$ &$\Z_{2l-1} \Z_{2l}$ & $\X_{2l-1}\X_{2l}$ & $\Y_{2l-1}\Y_{2l}$ \\ 
\hline
$\Ss_{l,0}$&$\frac{1}{4}$ &$\frac{1}{4}$ & $\frac{1}{4}$ & $-\frac{1}{4}$ \\ 
\hline
$\Ss_{l,1}$& $\frac{1}{4}$ &$\frac{1}{4}$ & $-\frac{1}{4}$ & $\frac{1}{4}$ \\ 
\hline
$\Ss_{l,2}$&$\frac{1}{4}$ & $-\frac{1}{4}$ & $\frac{1}{4}$  & $\frac{1}{4}$ \\ 
\hline
$\Ss_{l,3}$ & $\frac{1}{4}$ &$-\frac{1}{4}$ & $-\frac{1}{4}$  & $-\frac{1}{4}$ \\ 
\hline
\end{tabular}\\[10pt]
\begin{tabular}{ |c | c | c | c | c |} 
\hline
 & $\quad \idd \quad$ &$\Z_{2l} \Z_{2l+1}$ & $\X_{2l}\X_{2l+1}$ & $\Y_{2l}\Y_{2l+1}$ \\ [0.6ex]
\hline
$\T_{l,0}$&$\frac{1}{4}$ &$\frac{1}{4}$ & $\frac{1}{4}$ & $-\frac{1}{4}$ \\ 
\hline
$\T_{l,1}$& $\frac{1}{4}$ &$\frac{1}{4}$ & $-\frac{1}{4}$ & $\frac{1}{4}$ \\ 
\hline
$\T_{l,2}$&$\frac{1}{4}$ & $-\frac{1}{4}$ & $\frac{1}{4}$  & $\frac{1}{4}$ \\ 
\hline
$\T_{l,3}$ & $\frac{1}{4}$ &$-\frac{1}{4}$ & $-\frac{1}{4}$  & $-\frac{1}{4}$ \\ 
\hline
\end{tabular}
\caption{\label{tablecors} Elements of the table give correlation $\bra{\psi}C\otimes R\ket{\psi}$ where $C$ is the operator labelling the column and $R$ the operator labelling the row.}
\end{table}
To generate our self-testing correlations we use the same strategy as Lemma 2 for the inputs $\vx$ and $\vz$. The two new measurements for Alice $x=\lozenge, \blacklozenge$ correspond to Bell state measurements between successive pairs of qubits of her system, where the Bell state measurements for the input $\blacklozenge$ are shifted with respect to those for $\lozenge$ (see Fig.\ \ref{bsmfig}). Specifically,
\begin{align}
\Pi_{\va,\lozenge}&=\bigotimes_{l=1}^{\lfloor{\frac{n}{2}}\rfloor} \proj{\Psi_{a_i}}^{\rA_{2l-1}\rA_{2l}} \\
\Pi_{\va,\blacklozenge}&=\bigotimes_{l=1}^{\lfloor{\frac{n-1}{2}}\rfloor} \proj{\Psi_{a_i}}^{\rA_{2l}\rA_{2l+1}},
\end{align}
where $\{\ket{\Psi_{0}},\ket{\Psi_{1}},\ket{\Psi_{2}},\ket{\Psi_{3}}\}=\{\ket{\Phi^{\tx{+}}},\ket{\Phi^{\tx{-}}},\ket{\Psi^{\tx{+}}},\ket{\Psi^{\tx{-}}}\}$. With this choice, the correlations are given by Table \ref{tablecors}, which follow from the correlations of the four Bell states. We are now ready for our final self-testing lemma (see appendix \ref{bsmproof}).
\begin{tcolorbox}
\begin{lemma}\label{BSMlemma}
Let the state $\ket{\psi}\in\mathcal{H}_{\rC}\otimes\mathcal{H}_{\rA}$ and $\;\pm1\;$ outcome observables $\X^{\rC},\Y^{\rC},\Z^{\rC},\D_{\tx{zx}}^{\rA}, \E_{\tx{zx}}^{\rA},\D_{\tx{xy}}^{\rA}, \E_{\tx{xy}}^{\rA},\D_{\tx{zy}}^{\rA}, \E_{\tx{zy}}^{\rA}$ satisfy the conditions of Lemma \ref{lemma2} so that $\ket{\xi}$ has the form \eqref{junk2}. Furthermore, let projectors $\Ss_{l,a^*}$ and $\T_{l,a^*}$ satisfy the correlations given in Tables \ref{bsmtab1} for all $l$. Then $\ket{\xi}$ has the form
\begin{equation}\label{xiBSM2}
\ket{\xi} = \ket{\xi_{0}}\tp\ket{0\dots 0} + \ket{\xi_{1}}\tp\ket{1\dots 1}.
\end{equation}
\end{lemma}
\end{tcolorbox}
\noindent Note that $\ket{\xi}$ now has the form of definition \ref{defstcomplex} as desired. 

\subsection{Noise robustness}\label{stnoise}
It is important to study the noise robustness of Lemmas 1-3 as it is impossible to achieve perfect self-testing correlations in practice. In the same way as related works \cite{Col,McK2,Wu,CN}, Lemma \ref{lemma1} and Lemma \ref{lemma2} can be made noise robust. In appendix \ref{lemma1Rob} we show how precise robustness bounds can be estimated for Lemma 1. For instance, if we have a non-maximal value $\bra{\psi}\mathcal{B}\ket{\psi}=6\sqrt{2}-\epsilon$, equation \eqref{ststate} from Lemma \ref{lemma1} becomes
\begin{equation*}\label{qbitrobust}
\left\|U\left(\ket{\psi}^{\rC\rA}\otimes\ket{00}\right) -  \ket{\xi}^{\rC\rC''\rA\rA''}\tp \ket{\Phi^{\tx{+}}}^{\rC'\rA'}\right\| \leq c \sqrt{\epsilon},
\end{equation*}
where $c=55+36\sqrt{2}$. Similar statements can be derived for equations \eqref{stx} -- \eqref{sty}. For robust statements of Lemma \ref{lemma2}, we point the reader to \cite{Col} where the same techniques can be applied to our results to obtain polynomial robustness bounds; we do not elaborate further here since such calculations are based on well established methods and are not particularly enlightening. Concerning Lemma 3, we show that given a noise robust Lemma \ref{lemma2}, one can extend this to a robust version of Lemma \ref{BSMlemma} (see appendix \ref{lemma3Rob}). These robustness statements will become relevant later in order to make the entanglement certification protocols of Section \ref{scenariosection} tolerant to experimental noise.

\begin{figure*}
\includegraphics{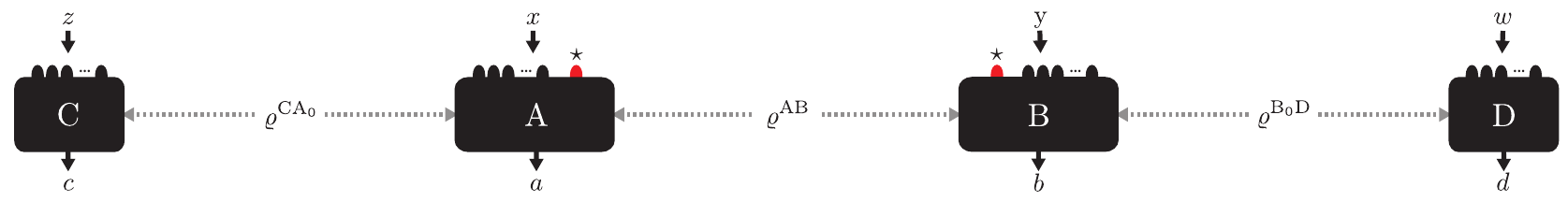}
\caption{\label{discenario} Device-independent scenario for our entanglement certification protocol. The correlations $p(c,a,b,d|z,x,y,w)$ are checked for (i) maximal violation of a Bell inequality in each of the marginal distributions $p(c,a|z,x)$, $p(b,d|y,w)$ which via self-testing certifies that the states $\varrho^{\rC\rA_0}$, $\varrho^{\rB_0\rD}$ are maximally entangled and that the measurements of Charlie and Daisy are Pauli measurements, and (ii) violation of an additional inequality $\mathcal{I}[p(c,a,b,d|z,x=\star,y=\star,w)]$ where Alice and Bob perform the measurements $x=\star$, $y=\star$, which certifies the entanglemt of $\varrho^{\rA\rB}$ given (i) is satisfied}. 
\end{figure*}

\section{Device-independent entanglement certification}\label{scenariosection}
In this section we show how to make use of the preceding self-testing results to construct device-independent entanglement certification protocols for all bipartite entangled quantum states. The precise scenario that we consider is a quantum network featuring three bipartite states: $\varrho^{\rA\rB}$ shared between Alice and Bob, and two auxiliary states $\varrho^{\rC\rA_0}$ and $\varrho^{\rB_0\rD}$ shared between Charlie and Alice, and Bob and Daisy respectively. Denoting the set of linear operators on Hilbert space $\mathcal{H}$ by $\mathcal{B}(\mathcal{H})$ we have $\varrho^{\rA\rB}\in\mathcal{B}(\mathcal{H}_{\rA}\otimes\mathcal{H}_{\rB})$, $\varrho^{\rC\rA_0}\in\mathcal{B}(\mathcal{H}_{\rC}\otimes\mathcal{H}_{\rA_0})$ and $\varrho^{B_0D}\in\mathcal{B}(\mathcal{H}_{\rB_0}\otimes\mathcal{H}_{\rD})$. We are interested in certifying the entanglement of the state $\varrho^{\rA\rB}$ when placed in a line network (see Fig.\ \ref{discenario}) featuring the auxillary states $\varrho^{\rC\rA_0}$ and $\varrho^{\rB_0\rD}$. In such a network, the correlations $\{p(c,a,b,d\vert z,x,y,w)\}$ are given by: 
\begin{align}\label{correlations}
&p(c,a,b,d\vert z,x,y,w)=\\ 
&\Tr\left[\M^\rC_{c|z}\otimes \M^{\rA_0\rA}_{a|x} \otimes \M^{\rB \rB_0}_{b|y} \otimes \M^{\rD}_{d|w} \;\varrho^{\rC\rA_0}\otimes\varrho^{\rA\rB}\otimes \varrho^{\rB_0\rD}\right],\nonumber
\end{align}
where the $\M_{i|j}$ are the local measurement operators for each party. In the device-independent scenario, one only has access to the observed correlations $p(c,a,b,d\vert z,x,y,w)$. Hence, a device-independent certification of the entanglement of $\varrho^{\rA\rB}$ is possible only if the observed correlations cannot be reproduced by \eqref{correlations} for any separable $\varrho^{\rA\rB}$. That is, one must show
\begin{align}\label{sepcors}
&p(c,a,b,d\vert z,x,y,w)\neq\\
&\Tr\left[\M'^\rC_{c|z}\otimes \M'^{\rA_0\rA}_{a|x} \otimes \M'^{\rB \rB_0}_{b|y} \otimes \M'^{\rD}_{d|w} \;\varrho'^{\rC\rA_0}\otimes\varrho_{\tx{SEP}}^{\rA\rB}\otimes \varrho'^{\rB_0\rD}\right] \nonumber
\end{align}
for any choice of separable $\varrho_{\tx{SEP}}^{\rA\rB}$, and any local measurement operators $\M'_{i\vert j}$ and auxillary states $\varrho'^{\rC\rA_0}$ and $\varrho'^{\rB_0\rD}$. 
Note that the auxiliary states may be entangled and that since we impose no constraints on the dimension of the auxiliary systems in \eqref{sepcors}, we may purify them and take all measurements to be projective without loss of generality. 

As we work in the device-independent scenario, all devices are treated as black boxes that process classical information. The precise assumptions we then make about the experiment are as follows.
\begin{enumerate}
\item States and measurements are described by quantum mechanics
\item The rounds of the experiment are independent and identically distributed (i.i.d.)
\item The network of Fig.\ \ref{discenario} correctly describes the experimental setup
\end{enumerate}
The first two of these assumptions are standard in device-independent studies (ideally one would like to drop the second assumption, see \cite{EA,DIprot} for some recent progress) . The last assumption is required so that we may write our probabilities in the form \eqref{correlations}. Physically this assumption means that one is able to prepare the three states independently and that they are trusted to interact in the way described by the network of Fig.\ \ref{discenario} (for example the state $\varrho^{\rC\rA_0}$ should only interact with Charlie and Alice and not Bob or Daisy). 

\subsection{Certification protocols}\label{protocolsectionintro}
\noindent We now present our entanglement certification protocols. These can be seen as a device-independent extension of the measurement device-independent entanglement witnesses (MDIEWs) presented in previous works \cite{Buscemi,MDIEW1,MDIEW2}. There, measurement devices are treated as black boxes, however inputs are given as a set of known informationally complete quantum states (in contrast to using classical variables as inputs). Then, an entanglement certification protocol can be built for every entangled state starting from an entanglement witness for the state. However, since this scheme requires a set of trusted input quantum states it is only partially device-independent. To see how these protocols can be made fully device-independent (i.e. how to remove the trust on the input states) consider that in the network of Fig.\ \ref{discenario} the auxiliary states are given by maximally entangled states and that the complete set of projectors for Charlie's  (resp.\ Daisy's) measurements form an informationally complete set. This can in fact be certified device-independently using the self-testing protocols of the first part of the paper (see Lemma 1 and Lemma 3). With this, the states that Alice (Bob) receives in the Hilbert space $\mathcal{H}_{\rA_0}$ ($\mathcal{H}_{\rB_0}$) conditioned on the different inputs and outputs of Charlie (Daisy) also form an informationally complete set. By interpreting these states as the inputs in a MDIEW protocol, one is essentially in the MDIEW scenario and the same techniques can be applied. Here, one has to be a bit careful due to the issue of transposition encountered in the self-testing sections, which we deal with in the appendix.

We now formalise this intuition and move to the main result of this section. \\ [8pt]
\emph{The entanglement of all bipartite entangled states can be certified device-independently in the network of Fig.\ \ref{discenario}.} \\[8pt]
\noindent In order to show this, we give an explicit family of entanglement certification protocols. The protocols we consider have the same structure for all states and are summarised as follows:\\
\begin{tcolorbox}
$\makebox[\linewidth]{\bf Entanglement certification protocol}$\\[3pt]
(i) \textbf{[generation of correlations]} The parties perform local measurements on their subsystems to obtain the correlations $p(c, a, b, d|z, x, y, w)$. \\ [4pt]
(ii) The following is then verified: \\[2pt]
\textbf{[self-testing]} The marginal distributions $p(c,a\vert z,x)$ and $p(b,d\vert y,w)$ maximally violate a Bell inequality that certifies that the auxiliary states each contain a maximally entangled state and that Charlie and Daisy each perform Pauli measurements on their subsystems. \\[2pt]
\textbf{[entanglement certification]} The correlations violate an additional inequality $\mathcal{I}(p(c, a, b, d|z, x, y, w)\geq 0$ that certifies $\varrho^{\rA\rB}$ is entangled.
\end{tcolorbox}

For now, we have the unrealistic requirement that we have a maximum violation of a Bell inequality in step (ii). This can be weakened to allow for some noise on the statistics, which we discuss in section \ref{dinoise}. We now describe in detail the above protocol, starting with the case of two-qubit states. 

\subsection{Entanglement certification of all two-qubit entangled states}
We start by defining the scenario in which we work. Charlie and Daisy both have a choice of three measurements $z,w=1,2,3$ and Alice and Bob both have a choice of seven inputs $x,y=1,2,3,4,5,6,\star$.  All outputs are $\pm1$ valued. 

{\emph{(i) Generation of correlations}}--- To generate the correlations in step (i) of the protocol, the parties chose $\varrho^{\rC\rA_0}=\varrho^{\rB_0\rD}=\proj{\Phi^{\tx{+}}}$. Measurements for inputs $z=1,2,3$ and $x=1,\cdots,6$ for Charlie and Alice should be chosen so that the conditions of Lemma 1 are satisfied, i.e. given by the qubit observables
\begin{align}\label{qubitm}
\sigma_{\tx{z}},\, \sigma_{\tx{x}},\, \sigma_{\tx{y}} \quad &z=1,2,3 \\  \label{qubitm2}
\frac{\sigma_{\tx{z}}\pm\sigma_{\tx{x}}}{\sqrt{2}}, \, \frac{\sigma_{\tx{z}}\pm\sigma_{\tx{y}}}{\sqrt{2}}, \, \frac{\sigma_{\tx{x}}\pm\sigma_{\tx{y}}}{\sqrt{2}} \quad &x=1,\cdots 6
\end{align}
acting on the $\mathcal{H}_{\rC}$ and $\mathcal{H}_{\rA_0}$ spaces respectively. Measurements for Daisy and Bob are defined analogously. Lastly, the measurement operators for inputs $x=\star$, $y=\star$ are projections onto the maximally entangled state: 
\begin{align}\label{qubitm3}
\M^{\rA\rA_0}_{+\vert\star}=\M^{\rB_0\rB}_{+\vert\star}=\proj{\Phi^{\tx{+}}}.
\end{align}

{\emph{(ii) Self-testing}}--- Our next step is to define the Bell inequality used in step (ii) of the protocol. Here we focus on Charlie and Alice; the Bell inequality used by Daisy and Bob is the same. The inequality we consider is constructed by combining three CHSH Bell inequalities \cite{CHSH}. Define the expectation value for inputs $z$, $x$ as
\begin{align}
E_{z,x}=\sum_{c,a=\pm1}c\cdot a \;p(c,a|z,x).
\end{align}
We then define the triple CHSH Bell inequality
\begin{multline}\label{chainedCHSH}
\mathcal{J}= E_{1,1}+E_{1,2}+E_{2,1}-E_{2,2}  \\
+E_{1,3}+E_{1,4}-E_{3,3}+E_{3,4}  \\
+E_{2,5}+E_{2,6}-E_{3,5}+E_{3,6}.
\end{multline}
Note that each line in the above is a CHSH inequality, and each of Charlie's inputs appears in two of the lines, and that at this stage the inputs $x,y=\star$ remain unused. Using the states and measurements above one finds $\mathcal{J}=6\sqrt{2}$. Via Lemma 1, this provides a self-test of the auxiliary states and measurements of Charlie and Daisy  defined in step (i), up to local transposition. 

\emph{Entanglement certification}--- Our next task is to construct the inequality used in the final step of the protocol. The inequality is constructed from an entanglement witness $\mathcal{W}$ for the state $\varrho^{\rA\rB}$. We thus have $\Tr[\mathcal{W} \sigma]\geq 0$ for all separable states $\sigma$ and $\Tr[\mathcal{W} \varrho^{\rA\rB}]< 0$. Consider the projectors $\pi_{c|z}=\frac{1}{2}[\openone+c\,\sigma_{z}]$ with $c=\pm1$ and $z=1,2,3$, that is, projectors onto the plus and minus eigenspaces of the Pauli operators. Since these form an (over-complete) basis of the set of Hermitian matrices, any entanglement witness may be decomposed as 
\begin{equation}
\label{witness}
\mathcal{W} = \sum_{cdzw}\omega_{cd}^{zw}\,\pi_{c\vert z}\otimes \pi_{d\vert w}.
\end{equation}
To define our inequality, we make use of the additional inputs for both Alice and Bob $x=\star$ and $y=\star$. The inequality is then given by
\begin{equation}
\label{ineq}
\mathcal{I}=\sum_{cdzw}\omega_{cd}^{zw}\;p(c,+,+,d\vert z,x=\star,y=\star,w)\geq 0
\end{equation}
and is satisfied for all separable states but violated using $\varrho^{\rA\rB}$. 

We first show that one can achieve $\mathcal{I}<0$ for entangled $\varrho^{\rA\rB}$. Using the states and measurements defined above one has 
\begin{align}\label{pproof1}
&p(c,+,+,d\vert z,x=\star,y=\star,w)= \\ 
&\Tr\left[\pi_{c|z}\sotimes\proj{\Phi^{\tx{+}}}\sotimes\proj{\Phi^{\tx{+}}}\sotimes\pi_{d|w}\, \proj{\Phi^{\tx{+}}}\sotimes \varrho^{\rA\rB}\sotimes\proj{\Phi^{\tx{+}}}\right] \nonumber\\
=&\frac{1}{4}\tr\left[\proj{\Phi^{\tx{+}}}\sotimes\proj{\Phi^{\tx{+}}}\, \pi_{c|z}^T\sotimes \varrho^{\rA\rB} \sotimes \pi_{d|w}^T\right] \\
=&\frac{1}{16}\tr\left[\pi_{c|z}\otimes \pi_{d|w} \,\varrho^{\rA\rB} \right], \label{pproof2}
\end{align}
where we have used $\tr_{\rA}[\proj{\Phi^{\tx{+}}}\, \pi^{\rA}_{i|j}\sotimes \idd]=\frac{1}{2}\pi_{i|j}^T$ in the third and fourth line. One thus has 
\begin{align}\label{viol}
\mathcal{I}&=\frac{1}{16}\sum_{czdw}\omega_{cd}^{zw}\tr[\pi_{c|z}\otimes \pi_{d|w} \,\varrho^{\rA\rB}] \\
\mathcal{I}&=\frac{1}{16}\Tr[\mathcal{W}\varrho^{\rA\rB} ]<0, \label{viol2}
\end{align}
which follows from the fact that $\mathcal{W}$ is an entanglement witness for the state.

We now consider the case in which $\varrho_{\rA\rB}$ is separable. In general, if the self-testing part of the protocol is satisfied then one can show that 
\begin{align}\label{posmap}
\mathcal{I}=\Tr\left[\mathcal{W}\,\Lambda(\varrho^{\rA\rB})\right],
\end{align}
where $\Lambda$ is a local, positive map on separable quantum states (see appendix \ref{entcertqubits} for details). Hence $\Lambda(\varrho^{\rA\rB})$ is a separable state and $\mathcal{I}\geq0$. A crucial observation in the proof of the above is that although the measurements for Charlie and Daisy are only certified via self-testing up to a possible transposition, this uncertainty can be mapped to possible local transpositions on the state $\varrho^{\rA\rB}$. Since local transpositions map separable states to separable states, this ensures that a false-positive certification of entanglement does not occur.

\subsection{Entanglement certification of high dimensional states}
The previous protocol for two-qubit states can be applied in parallel to construct entanglement certification protocols for bipartite states of any dimension. In the following we construct protocols for states of local dimension $2^n$ where $n=2,3,\cdots$. Since a state of local dimension $d$ can be seen as a particular case of a state of dimension $2^n$ for some $n\geq \log_2 d$ this implies a protocol for any dimension. 

The scenario we consider is as follows. Charlie and Daisy each have $3^n$ inputs, given by the vectors $\vz=(z_1,\cdots,z_n)$ and $\vw=(w_1,\cdots,w_n)$ with $z_i,w_i=1,2,3$, each with $2^n$ outcomes given by $\vc=(c_1,\cdots,c_n)$ and $\vd=(d_1,\cdots,d_n)$ with $c_i,d_i=\pm1$. Alice and Bob each have $6^{n}$ inputs given by the vectors $\vx=(x_1,\cdots,x_n)$, $\vy=(y_1,\cdots,y_n)$ with $x_i,y_i=1,\cdots, 6$, with outcomes $\va=(a_1,\cdots,a_n)$, $\vb=(b_1,\cdots,b_n)$ with $a_i,b_i=\pm1$. Further to this Alice and Bob have each two additional inputs $x=\lozenge,\blacklozenge$ and $y=\lozenge,\blacklozenge$ with $4^{\lfloor\frac{n}{2}\rfloor}$ and $4^{\lfloor\frac{n-1}{2}\rfloor}$ outputs respectively (as in Lemma 3), and inputs $x=\star$ and $y=\star$ with outputs $a=\pm 1$, $b=\pm 1$ (to be used in step (iii) of the protocol). 

\emph{(i) Generation of correlations}--- Since we will perform the previous protocol in parallel, the Hilbert spaces of the auxiliary systems are written as the tensor product of $n$ qubit spaces: $\mathcal{H}_{\rC}=\otimes_i \mathcal{H}_{\rC_i}$,  $\mathcal{H}_{\rA_0}=\otimes_i \mathcal{H}_{\rA_{0i}}$ (and similarly for Daisy, Bob). The auxiliary states are then $n$-fold tensors of maximally entangled states on each two-qubit subspace:
\begin{align}
\varrho^{\rC\rA_0}=\otimes_{i=1}^{n}\proj{\Phi^{\tx{+}}}^{\rC_i\rA_{0i}}; \quad \varrho^{\rB_{0}\rD}=\otimes_{i=1}^{n}\proj{\Phi^{\tx{+}}}^{\rB_{0i}\rD_{i}}. \nonumber
\end{align}
Measurements are a parallel version of the measurements \eqref{qubitm}, \eqref{qubitm2}, i.e. they are given by $n$-fold tensor products of the measurements \eqref{qubitm}, \eqref{qubitm2}, acting on each maximally entangled state. For example $z_i=1,2,3$ corresponds to a measurement of $\sigma_{\tx{z}}, \sigma_{\tx{x}}, \sigma_{\tx{y}}$ on the $i^{th}$ subsystem of Charlie with outcome $c_{i}$. As before, the measurements $\M_{+\vert\star}$ are projections onto the maximally entangled state: 
\begin{align}
\M^{\rA\rA_0}_{+\vert\star}=\M^{\rB_0\rB}_{+\vert\star}=\proj{\Phi^{\tx{+}}}
\end{align}
where here $\ket{\Phi^{\tx{+}}}=\frac{1}{\sqrt{2^n}}\sum_{i}\ket{ii}\in\mathcal{H}_{\rC}\otimes\mathcal{H}_{\rA_0}$. Finally, the measurements for the inputs $x,y=\lozenge,\blacklozenge$ are chosen to be tensor products of Bell state measurements between successive pairs of qubits of the local subsystems of Alice and Bob, and where the Bell state measurements for the input $\lozenge$ are shifted with respect to those for $\blacklozenge$ (see Fig.\ \ref{bsmfig} and Section \ref{bsmsection} for more details). 

\emph{(ii) Self-testing}--- The Bell inequality is now a parallel version of \eqref{chainedCHSH} (again we just describe the inequality for Charlie and Alice). Define the average expectation value for the bits $c_i$, $a_i$ given $z_i=z$, $x_i=x$ as
\begin{align}
E^{i}_{z,x}=\frac{1}{3^{n-1}6^{n-1}}\sum\limits_{\substack{{\vz}\vert z_i=z\\ \vx\vert x_i=x}}\sum_{\vc,\va} c_i\cdot a_i\,p(\vc,\va\vert \vz,\vx). 
\end{align}
For each $i$, we now have the triple CHSH Bell inequality: 
\begin{multline}\label{chainedCHSHparallel}
\mathcal{J}_i= E^i_{1,1}+E^i_{1,2}+E^i_{2,1}-E^i_{2,2}  \\
+E^i_{1,3}+E^i_{1,4}-E^i_{3,3}+E^i_{3,4} \\
+E^i_{2,5}+E^i_{2,6}-E^i_{3,5}+E^i_{3,6}.
\end{multline}
For the entanglement certification protocol we require maximum violation of each of these inequalities, i.e.\
\begin{align}
\sum_{i=1}^n\mathcal{J}_i=n\cdot 6\sqrt{2}.
\end{align}
We further require that the measurements $x,y=\lozenge,\blacklozenge$ correctly reproduce the Bell state measurement correlations given in tables \ref{bsmtab1}, which is achieved by our chosen measurement strategy and detailed in section \ref{bsmsection}. With these conditions met, we may apply Lemma 3 and move on to the entanglement certification of $\varrho^{\rA\rB}$.

\emph{(iii) Entanglement certification}--- Similarly to \eqref{witness}, we may decompose an entanglement witness for $\varrho^{\rA\rB}\in\otimes_i[\mathcal{H}_{\rA_i}\otimes\mathcal{H}_{\rB_i}]$ using tensor products of Pauli projectors as an (over-complete) basis: 
\begin{align}\label{witgen}
\mathcal{W}= \sum_{\vc,\vd,\vz,\vw}\omega_{\vc\vd}^{\vz\vw}\otimes_i \left[\pi_{c_i|z_i}^{\rA_i}\otimes \pi_{d_i|w_i}^{\rB_i}\right]. 
\end{align}
The inequality that is used to certify entanglement is then
\begin{align}\label{jaime}
\mathcal{I}=\sum_{\vc,\vd,\vz,\vw}\omega_{\vc\vd}^{\vz\vw} p(\vc,+,+,\vd\vert \vz,x=\star,y=\star,\vw)\geq 0,
\end{align}
which for separable states gives
\begin{align}\label{posmap2}
\mathcal{I}=\Tr\left[\mathcal{W}\,\Lambda(\varrho^{\rA\rB})\right]\geq 0,
\end{align}
where $\Lambda$ is again a local positive map on separable states (see appendix \ref{entcertgen} for a full proof). Note here that simply using two-qubit strategy in parallel (i.e. using Lemma \ref{lemma2}) without the additional Bell state measurements for inputs $x,y=\lozenge,\blacklozenge$ would lead to problems. This is because the measurements for Charlie and Daisy would be certified only up to possible flipping of any number of their $n$ $\sigma_{\tx{y}}$ measurements. When mapping this uncertainty to the state $\varrho^{\rA\rB}$, this corresponds to possible local transposition on \emph{part of a local subsystem} of $\varrho^{\rA\rB}$, which may map separable states to unphysical (non-positive) states. Hence, the additional Bell state measurements ensure that either none or all $\sigma_{\tx{y}}$ measurements are flipped, corresponding to a transposition of the entire local subsystem of $\varrho^{\rA\rB}$ so that the map $\Lambda$ is positive on separable states.  

Finally, we show that $\mathcal{I}$ is violated by $\varrho^{\rA\rB}$. Using the measurement strategy above and that  $\tr_{\rA}[\proj{\Phi^{\tx{+}}}\, \pi^{\rA}_{i|j}\sotimes \idd]=\frac{1}{d}\pi_{i|j}^T$ for the maximally entangled state of dimension $d$, it is straightforward to show using the same technique as \eqref{pproof1} - \eqref{pproof2} that
\begin{align}
\mathcal{I}&=\frac{1}{d^4}\sum_{\vc,\vd,\vz,\vw}\omega_{\vc\vd}^{\vz\vw} \tr\left[ \otimes_i \left(\pi_{c_i|u_i}^{\rA_i}\otimes \pi_{d_i|w_i}^{\rB_i}\right) \varrho^{\rA\rB}\right] \\ \label{finalproof}
&=\frac{1}{d^4}\tr\left[\mathcal{W}\varrho^{\rA\rB}\right] <0,
\end{align}
thus certifying the entanglement of $\varrho^{\rA\rB}$. 

\subsection{Noise robust entanglement certification}\label{dinoise}

A natural question to ask is whether the above certification protocols can be extended to tolerate small amounts of experimental noise. Indeed, this can be achieved using robust versions of Lemmas \ref{lemma1} and \ref{BSMlemma}. The intuitive argument goes as follows. Imagine each of our probabilities differ from the ideal self-testing statistics by some small amount $\epsilon$. Then, the states that Alice and Bob receive from the auxiliary systems conditioned on Charlie's and Daisy's measurement outcomes should be close to eigenstates of products of Pauli operators. This implies that the analogous operator to $\mathcal{W}$ appearing in \eqref{finalproof} is close to the desired witness, which can be used to bound the maximum value of $\mathcal{I}$ for separable states to be
\begin{align}\label{robustI}
\mathcal{I}\geq -c(\epsilon)
\end{align}
for some positive function $c(\epsilon)$ such that $c(0)=0$. Unsurprisingly, this means that some weakly entangled states close to the separable set are no longer certified by the method. The amount of noise that can be tolerated by a typical state before it can no longer be certified depends on the optimality of the robustness bounds of the self-testing lemmas; given current techniques the noise tolerance is expected to be small. For a detailed proof and discussion of \eqref{robustI} see appendix \ref{REC}. For a specific analysis for the class of two-qubit Werner states, see Appendix \ref{isonoise}.

\section{Discussion and Conclusion} 
We have shown that all bipartite entangled quantum states are capable of producing correlations that cannot be obtained using separable states by placing them in a larger network of auxiliary states and using tools from self-testing and measurement device-independent entanglement witnesses. It is desirable to strengthen the self-testing part of our protocol; in particular, improved robustness bounds for self-testing would immediately translate into better noise-tolerance of our protocols. One would most likely be able to achieve this using the protocols presented in \cite{leash} where self-testing statements for Pauli observables are presented with a robustness scaling that is independent of $n$. Furthermore, the choice of measurements used for self-testing could be made much more efficient. In general, one needs $d^2$ linearly independent projectors to form an informationally complete set, however for local dimension $2^n$ we make use of an over-complete basis of $6^n$ projectors (coming from the tensor product of Pauli projectors), a difference that is exponential in $n$. Hence, a more efficient self-test of informationally complete sets of measurements would improve the efficiency of the protocol. Furthermore, given a particular state, one typically does not need the full set of projectors in order to write an entanglement witness for the state. It would therefore be interesting to study self-testing protocols that certify only those projectors that appear in a particular decomposition of an entanglement witness. 

Although we have focused on the task of entanglement certification, our technique can in principle be applied to other convex sets of quantum states other than the separable set where linear witnesses can also be used.  Due to the ambiguity of local unitaries and local transpositions in the self-testing part of our protocol, such sets would need to be closed under local unitary operations and local transpositions (as is the case for the separable set). For example, one could apply the same technique to certify entangled states with negative partial transpose. Finally, it would also be interesting to investigate the possibility of using our general technique for other device-independent tasks, for example using similar ideas to \cite{Lim,MDIQKD1,MDIQKD2} to construct device-independent quantum key distribution protocols, or to generalise our protocol for the certification of genuine multipartite entanglement.

\section{acknowledgements}
The authors are thankful for useful discussions to Paul Skrzypczyk, Nicolas Brunner, Marco T{\'u}lio Quintino, Flavien Hirsch, Thomas Vidick, Matteo Lostaglio,  Micha\l{} Oszmaniec and Alexia Salavrakos. This work was supported by the Ram\'on y Cajal fellowship, Spanish MINECO (QIBEQI FIS2016-80773-P and Severo Ochoa SEV-2015-0522), the AXA Chair in Quantum Information Science, Generalitat de Catalunya (CERCA Programme), Fundaci\'{o} Privada Cellex and ERC CoG QITBOX.

\onecolumngrid
\begin{appendix}
\section{Proof of Lemma 1}\label{lemma1proof}
In this section we prove Lemma \ref{lemma1} from the main text. Define the following operators:
\begin{align}
{\Z}^{\rA}_{\tx{z,x}} &= \frac{\D_{\tx{z,x}}^{\A} + \E_{\tx{z,x}}^{\A}}{\sqrt{2}},\quad {\X}^{\A}_{\tx{z,x}} = \frac{\D_{\tx{z,x}}^{\rA} - \E_{\tx{z,x}}^{\rA}}{\sqrt{2}}, \quad {\Z}^{\rA}_{\tx{z,y}} = \frac{\D_{\tx{z,y}}^{\rA} + \E_{\tx{z,y}}^{\rA}}{\sqrt{2}},\nonumber\\
{\Y}^{\rA}_{\tx{z,y}} &= \frac{\D_{\tx{z,y}}^{\rA} - \E_{\tx{z,y}}^{\rA}}{\sqrt{2}}, \quad {\X}^{\rA}_{\tx{x,y}} = \frac{\D_{\tx{x,y}}^{\rA} + \E_{\tx{x,y}}^{\rA}}{\sqrt{2}},\quad {\Y}^{\rA}_{\tx{x,y}} = \frac{\D_{\tx{x,y}}^{\rA} - \E_{\tx{x,y}}^{\rA}}{\sqrt{2}}. \label{defs}
\end{align}
From the \eqref{xyzcons1} -- \eqref{xyzcons3} we have  
\begin{equation}
{\Z}^{\rA}_{\tx{z,x}}\ket{\psi} = {\Z}^{\rA}_{\tx{z,y}}\ket{\psi} , \quad {\X}^{\rA}_{\tx{z,x}}\ket{\psi} = {\X}^{\rA}_{\tx{x,y}}\ket{\psi}, \quad {\Y}^{\rA}_{\tx{z,y}}\ket{\psi} = {\Y}^{\rA}_{\tx{x,y}}\ket{\psi}.
\end{equation}
Hence, defining
\begin{align}\label{a3}
{\Z}^{\rA}\equiv{\Z}^{\rA}_{\tx{z,x}}, \quad \X^{\rA}\equiv {\X}^{\rA}_{\tx{z,x}}, \quad \Y^{\rA}\equiv {\Y}^{\rA}_{\tx{z,y}}
\end{align}
we have from \eqref{xyzcons1} -- \eqref{ac} the conditions
\begin{align}\label{a4}
\Z^{\rC}\ket{\psi}= \Z^{\rA}\ket{\psi}, \quad \X^{\rC}\ket{\psi} =  \X^{\rA}&\ket{\psi}, \quad \Y^{\rC}\ket{\psi}= -\Y^{\rA}\ket{\psi},\\
\{\Z^{\rC},\X^{\rC}\}\ket{\psi} = 0, \quad \{\Z^{\rC},\Y^{\rC}\}\ket{\psi} &= 0, \quad \{\Y^{\rC},\X^{\rC}\}\ket{\psi}= 0, \quad  \\
\{\Z^{\rA},\X^{\rA}\}\ket{\psi} = 0, \quad \{\Z^{\rA},\Y^{\rA}\}\ket{\psi} &= 0, \quad \{\Y^{\rA},\X^{\rA}\}\ket{\psi}= 0.
\end{align}
Note that the operators $\Z^{\rA}$, $\X^{\rA}$, $\Y^{\rA}$ are not necessarily unitary. We may define the regularized versions of these operators $\hat{\Z}^{\rA}$, $\hat{\X}^{\rA}$, $\hat{\Y}^{\rA}$ which are obtained from the original operators by renormalising all eigenvalues to $\pm{1}$ and setting any zero eigenvalues to 1 (without changing the eigenvectors). Using standard techniques (for example see \cite{BampsS,SASA}) one can show that the regularized operators respect the same conditions, that is,
\begin{align}\label{qcons1}
\Z^{\rC}\ket{\psi}= \hat{\Z}^{\rA}\ket{\psi}, \quad \X^{\rC}\ket{\psi} =  \hat{\X}^{\rA}&\ket{\psi}, \quad \Y^{\rC}\ket{\psi}= -\hat{\Y}^{\rA}\ket{\psi},\\ \label{qcons2}
\{\Z^{\rC},\X^{\rC}\}\ket{\psi} = 0, \quad \{\Z^{\rC},\Y^{\rC}\}\ket{\psi} &= 0, \quad \{\Y^{\rC},\X^{\rC}\}\ket{\psi}= 0, \quad  \\ \label{qcons3}
\{\hat{\Z}^{\rA},\hat{\X}^{\rA}\}\ket{\psi} = 0, \quad \{\hat{\Z}^{\rA},\hat{\Y}^{\rA}\}\ket{\psi} &= 0, \quad \{\hat{\Y}^{\rA},\hat{\X}^{\rA}\}\ket{\psi}= 0.
\end{align}
Let us prove the first equality from \eqref{qcons1}, the other two being analogous. The following chain of equalities is satisfied
\begin{align}\label{regularize}
|| (\hat{\Z}^{\rA}-\Z^{\rA})\ket{\psi}|| &= || (\openone-(\hat{\Z}^{\dagger})^{\rA}\Z^{\rA})\ket{\psi}|| = || (\openone-|\Z^{\rA}|)\ket{\psi}||\\
&= || (\openone-|\Z^{\rC}\Z^{\rA}|)\ket{\psi}|| \leq || (\openone-\Z^{\rC}\Z^{\rA})\ket{\psi}|| = 0,
\end{align}
where the first equality comes from the fact that $(\hat{\Z}^{\dagger})^{\rA}$ is unitary, the second equality just uses the definition of ${\hat{\Z}}^{\rA}$. The third equality is equivalent to $|\Z^{\rC}\Z^{\rA}| = |\Z^{\rA}|$, which is correct because $\Z^{\rC}$ is unitary. The inequality is a consequence of $A \leq |A|$, and finally the last equality is the consequence of \eqref{a4}.

 We may now verify equations \eqref{ststate} to \eqref{junk} of Lemma \ref{lemma1} using the above conditions. The precise isometry that we use is shown in Fig.\ \ref{fig:STcircuit}. We first verify that the circuit acts correctly on the state $\ket{\psi}^{\rC\rA}$. Up to and including the second set of controlled gates the circuit is the well known SWAP circuit, and it is well known (see e.g. \cite{McKMS}) that this extracts the maximally entangled state in to the primed auxiliary systems. At this point our state is thus
\begin{align}\label{a10}
\ket{++}^{\rC''\rA''}\frac{\idd + \Z^\rC}{\sqrt{2}}\ket{\psi}^{\rC\rA}\otimes \ket{\Phi^{\tx{+}}}^{\rC'\rA'}.
\end{align}
Let us denote $\ket{\phi}^{\rC\rA} = \frac{1}{\sqrt{2}}[\idd + \Z^\rC]\ket{\psi}^{\rC\rA}$. The third pair of controlled gates evolves the system to
\begin{align}
\frac{1}{2}\left[\ket{00}^{\rC''\rA''}\ket{\phi}^{\rC\rA} + \ket{01}^{\rC''\rA''}i\hat{\Y}^{\rA}\hat{\X}^{\rA}\ket{\phi}^{\rC\rA} + \ket{10}^{\rC''\rA''}i\Y^{\rC}\X^{\rC}\ket{\phi}^{\rC\rA} - \ket{11}^{\rC''\rA''}\Y^{\rC}\X^{\rC}\hat{\Y}^{\rA}\hat{\X}^{\rA}\ket{\phi}^{\rC\rA}\right] \ket{\Phi^{\tx{+}}}^{\rC'\rA'}. \nonumber
\end{align}
From \eqref{qcons1} - \eqref{qcons3} it follows that $\hat{\Y}^{\rA}\hat{\X}^{\rA}\ket{\phi}^{\rC\rA} = \Y^{\rC}\X^{\rC}\ket{\phi}^{\rC\rA}$ and so
\begin{align}\label{a11}
\frac{1}{2}\left[\ket{00}^{\rC''\rA''}\ket{\phi}^{\rC\rA} + \ket{01}^{\rC''\rA''}i\Y^{\rC}\X^{\rC}\ket{\phi}^{\rC\rA} + \ket{10}^{\rC''\rA''}i\Y^{\rC}\X^{\rC}\ket{\phi}^{\rC\rA} + \ket{11}^{\rC''\rA''}\ket{\phi}^{\rC\rA}\right] \ket{\Phi^{\tx{+}}}^{\rC'\rA'}.
\end{align}
Finally the last two Hadamards lead to
\begin{align}
&\frac{1}{2\sqrt{2}}\left[\ket{00}^{\rC''\rA''}(\idd + i\Y^\rC\X^\rC)(\idd + \Z^\rC)\ket{\psi}^{\rC\rA} + \ket{11}^{\rC''\rA''}(\idd - i\Y^\rC\X^\rC) (\idd + \Z^\rC)\ket{\psi}^{\rC\rA}\right]\ket{\Phi^{\tx{+}}}^{\rC'\rA'} \\
&= \ket{\xi}^{\rC\rC''\rA\rA''}\otimes \ket{\Phi^{\tx{+}}}^{\rC'\rA'}
\end{align}
as claimed. 
Following the same method and using \eqref{qcons1} - \eqref{qcons3}, one easily verifies 
\begin{align}
U\left(\X^\rC\ket{\psi}^{\rC\rA}\otimes\ket{00}\right) =  \ket{\xi}^{\rC\rC''\rA\rA''}\tp \sigma_\tx{x}^{\rC'}\ket{\Phi^{\tx{+}}}^{\rC'\rA'}, \qquad U\left(\Z^\rC\ket{\psi}^{\rC\rA}\otimes\ket{00}\right) = \ket{\xi}^{\rC\rC''\rA\rA''}\tp\sigma_\tx{z}^{\rC'}\ket{\Phi^{\tx{+}}}^{\rC'\rA'}.
\end{align}
The case $\Y^\rC\ket{\psi}^{\rC\rA}\otimes\ket{00}$ is a bit more involved. After the second pair of controlled gates the state is transformed to 
\bn
\ket{++}^{\rC''\rA''} \frac{1}{\sqrt{2}}i\Y^\rC\X^\rC(\idd + \Z^\rC)\ket{\psi}^{\rC\rA}\sigma_{\tx{y}}^{\rC'}\ket{\Phi^{\tx{+}}}^{\rC'\rA'}.
\en
The third pair of controlled gates then transforms the state to
\bn
\frac{1}{4\sqrt{2}}\left[\ket{00}^{\rC''\rA''}i\Y^\rC\X^\rC\ket{\phi}^{\rC\rA} + \ket{01}^{\rC''\rA''}\ket{\phi}^{\rC\rA} + \ket{10}^{\rC''\rA''}\ket{\phi}^{\rC\rA} + \ket{11}^{\rC''\rA''}i\Y^\rC\X^\rC\ket{\phi}^{\rC\rA}\right]\sigma_{\tx{y}}^{\rC'}\ket{\Phi^{\tx{+}}}^{\rC'\rA'},
\en
which is simplified by two last Hadamards to
\begin{align}
&\frac{1}{2\sqrt{2}}\left[\ket{00}^{\rC''\rA''}(\idd + i\Y^\rC\X^\rC)(\idd + \Z^\rC)\ket{\psi}^{\rC\rA} - \ket{11}^{\rC''\rA''}(\idd - i\Y^\rC\X^\rC) (\idd + \Z^\rC)\ket{\psi}^{\rC\rA}\right]\sigma_\tx{y}^{\rC'}\ket{\Phi^{\tx{+}}}^{\rC'\rA'}.  \\
&= \sigma_{\tx{z}}^{\rC''} \ket{\xi}^{\rC\rC''\rA\rA''}\tp \sigma_\tx{y}^{C'}\ket{\Phi^{\tx{+}}}^{\rC'\rA'} 
\end{align}
This thus concludes the proof of Lemma \ref{lemma1}.

\section{Robust version of Lemma 1}\label{lemma1Rob}
 
Following the approach from \cite{YNS} and \cite{MKSS} we study how Lemma 1 is affected when the achieved Bell inequality (\ref{optCorr}) violation  is $6\sqrt{2}-\epsilon$. Looking at SOS decomposition \eqref{chshsos} one can see that each of the terms must be smaller or equal to $\sqrt{\epsilon}$, leading to:
\begin{align}\label{a4Rob}
\|(\Z^{\rC} - \Z^{\rA})\ket{\psi}\| \leq \sqrt{\epsilon}, \quad \|(\X^{\rC} -  \X^{\rA})\ket{\psi}\| &\leq \sqrt{\epsilon}, \quad \|(\Y^{\rC}+\Y^{\rA})\ket{\psi}\| \leq \sqrt{\epsilon},\\
\|(\Z^{\rC} - \hat{\Z}^{\rA})\ket{\psi}\| \leq 2\sqrt{\epsilon}, \quad \|(\X^{\rC} -  \hat{\X}^{\rA})\ket{\psi}\| &\leq 2\sqrt{\epsilon}, \quad \|(\Y^{\rC}+\hat{\Y}^{\rA})\ket{\psi}\| \leq 2\sqrt{\epsilon},\\
\|\{\Z^{\rC},\X^{\rC}\}\ket{\psi}\| \leq (4+4\sqrt{2})\sqrt{\epsilon}, \quad \|\{\Z^{\rC},\Y^{\rC}\}\ket{\psi}\| &\leq (6+6\sqrt{2})\sqrt{\epsilon}, \quad \|\{\Y^{\rC},\X^{\rC}\}\ket{\psi}\| \leq (8+8\sqrt{2})\sqrt{\epsilon}.
\end{align}
Let us first note that the error coming from the regularizing operators on Alice's side is
\begin{align*}
|| (\hat{\Z}^{\rA}-\Z^{\rA})\ket{\psi}|| &= || (\openone-(\hat{\Z}^{\dagger})^{\rA}\Z^{\rA})\ket{\psi}|| = || (\openone-|\Z^{\rA}|)\ket{\psi}||\\
&= || (\openone-|\Z^{\rC}\Z^{\rA}|)\ket{\psi}|| \leq || (\openone-\Z^{\rC}\Z^{\rA})\ket{\psi}|| = \sqrt{\epsilon},
\end{align*}
and similarly for $\hat{\X}^{\rA}$ and $\hat{\Y}^{\rA}$. Taking this into account inequalities in the second line follow from the corresponding inequalities in the first line and the triangle inequality $\|a + b\| \leq \|a\| + \|b\|$. 
The first inequality in the third line is obtained through the following chain of inequalities
\begin{align*} 
\|(\Z^{\rC}\X^{\rC} + \X^{\rC}\Z^{\rC})\ket{\psi}\| &\leq \| \Z^{\rC}(\X^{\rC}-\X^{\rA}) \ket{\psi}\| + \|(\Z^{\rC}\X^{\rA} + \X^{\rC}\Z^{\rA})\ket{\psi}\| + \|\X^{\rC}(\Z^{\rA}-\Z^{\rC})\ket{\psi}\| \\
&\leq \sqrt{\epsilon} + \|\X^{\rA}(\Z^{\rC}-\Z^{\rA})\ket{\psi}\| + \|(\Z^{\rA}\X^{\rA} + \X^{\rA}\Z^{\rA})\ket{\psi}\| + \|\Z^{\rA}(\X^{\rC}-\X^{\rA}) \ket{\psi}\| + \sqrt{\epsilon} \\
&\leq 2\sqrt{\epsilon} +  \frac{1}{\sqrt{2}}\|(\D_{\tx{z,x}}^{\rA} - \E_{\tx{z,x}}^{\rA})(\Z^{\rC}-\Z^{\rA})\ket{\psi}\| + \|(\Z^{\rA}\X^{\rA} + \X^{\rA}\Z^{\rA})\ket{\psi}\| + \frac{1}{\sqrt{2}}\|(\D_{\tx{z,x}}^{\A} + \E_{\tx{z,x}}^{\A})(\X^{\rC}-\X^{\rA}) \ket{\psi}\| \\
&\leq (2+2\sqrt{2})\sqrt{\epsilon} + \|\Z^{\rA}(\X^{\rA}-\hat{\X}^{\rA})\ket{\psi}\| + 
\|\hat{X}^{\rA}(\Z^{\rA}-\hat{\Z}^{\rA})\ket{\psi}\| + \|(\hat{\Z}^{\rA}\hat{\X}^{\rA} + \hat{\X}^{\rA}\hat{\Z}^{\rA})\ket{\psi}\| + \\ & \qquad \qquad + \|\X^{\rA}(\Z^{\rA}-\hat{\Z}^{\rA})\ket{\psi}\| + \|\hat{\Z}^{\rA}(\X^{\rA}-\hat{\X}^{\rA})\ket{\psi}\|\\
&\leq (4+4\sqrt{2})\sqrt{\epsilon}
\end{align*}
Note that if the violation of Bell inequality is $6\sqrt{2}-\epsilon$ not all terms from the first line of \eqref{a4Rob} can simultaneously be equal to $\sqrt{\epsilon}$, but for our purposes a tight multiplicative factor is not of primary interest. The second and the third inequality from the third line of \eqref{a4Rob} are derived in an analogous manner as the first one, with the additional factors coming from the convention used in \eqref{a3} which leads to:
\begin{equation*}
\|(\Z^{\rA} - \Z_{\tx{z,y}}^{\rA})\ket{\psi}\| \leq \sqrt{\epsilon}, \quad \|(\X^{\rA} -  \X_{\tx{x,y}}^{\rA})\ket{\psi}\| \leq \sqrt{\epsilon}, \quad \|(\Y^{\rA}-\Y_{\tx{x,y}}^{\rA})\ket{\psi}\| \leq \sqrt{\epsilon}.
\end{equation*}
To check the error accumulated when obtaining the final statement from Lemma 1 we will repeatedly use the triangle inequality and bounds from \eqref{a4Rob}. To get \eqref{a10} the first inequality from the second line of \eqref{a4Rob} has to be used four times, the second one is used twice and the anticommuting bound from the third line of \eqref{a4Rob} has to be used once. To obtain \eqref{a11} the second and the third inequality from the second line and all three inequalities from the third line of \eqref{a4Rob} are each used twice. All together these bounds imply
\begin{equation*}
\|U\left(\ket{\psi}^{\rC\rA}\otimes\ket{00}\right) -  \ket{\xi}^{\rC\rC''\rA\rA''}\tp \ket{\Phi^{\tx{+}}}^{\rC'\rA'}\| \leq (55 + 36\sqrt{2})\sqrt{\epsilon}.
\end{equation*}
A similar asymptotic bounds can be obtained for the robust versions of Eqs. \eqref{stx}, \eqref{sty} and \eqref{stz}, the only difference being in the number of times each of the inequalities from \eqref{a4Rob} have to be used.


\section{Proof of Lemma 2}\label{lemma2proof}
The proof of Lemma 2 is split into two parts. The first part proves the necessary self-testing relations between the state and measurements needed to construct the self-testing circuit. The second part verifies that the circuit acts as claimed. 
\subsection{Self-testing relations}
Here we follow closely the proof of \cite{ColS}, adapting it the allow for additional $\sigma_{\tx{y}}$ measurements. We first define the following sets of operators:
\begin{align}\label{zks}
\{\Z_{i}^{(k)}\}_{k}=\{\Op_{i|\vz} \vert z_i=1\} ; \quad \{\X_{i}^{(k)}\}_{k}=\{\Op_{i|\vz} \vert z_i=2\} ;\quad \{\Y_{i}^{(k)}\}_{k}=\{\Op_{i|\vz} \vert z_i=3\},
\end{align}
for $k=1,\cdots, 3^{n-1}$ and ordered according to some relation $\vz<\vz'$. Similarly we define 
\begin{align}
\{\D_{\tx{zx},i}^{(l)}\}_l=\{\Pp_{i|\vx}\vert x_i=1\}; \quad  \{\E_{\tx{zx},i}^{(l)}\}_l&=\{\Pp_{i|\vx}\vert x_i=2\};\quad  \{\D_{\tx{zy},i}^{(l)}\}_l=\{\Pp_{i|\vx}\vert x_i=3\}, \\
  \{\E_{\tx{zy},i}^{(l)}\}_l=\{\Pp_{i|\vx}\vert x_i=4\}; \quad \{\D_{\tx{xy},i}^{(l)}\}_l&=\{\Pp_{i|\vx}\vert x_i=5\}; \quad  \{\E_{\tx{xy},i}^{(l)}\}_l=\{\Pp_{i|\vx}\vert x_i=6\}. 
\end{align}
for $l=1,\cdots,6^{n-1}$ ordered according to some relation $\vx<\vx'$. Averaging over these sets we thus obtain the operators in equations \eqref{zop} - \eqref{exop}. We may now write 
\begin{align}
\bra{\psi}\mathcal{B}_i\ket{\psi}=&\frac{1}{3^{n-1}6^{n-1}}\sum_{k,l}\bra{\psi}\bigg[\Z_i^{(k)}(\D^{(l)}_{\tx{zx},i}+\E^{(l)}_{\tx{zx},i})+\X^{(k)}_i(\D^{(l)}_{\tx{zx},i}-\E^{(l)}_{\tx{zx},i}) + \Z^{(k)}_i(\D^{(l)}_{\tx{zy},i}+\E^{(l)}_{\tx{zy},i})\nonumber\\&\quad -\Y^{(k)}_i(\D^{(l)}_{\tx{zy},i}-\E^{(l)}_{\tx{zy},i}) +\X^{(k)}_i(\D^{(l)}_{\tx{xy},i}+\E^{(l)}_{\tx{xy},i})-\Y^{(k)}_i(\D^{(l)}_{\tx{xy},i}-\E^{(l)}_{\tx{xy},i})\bigg]\ket{\psi}=6\sqrt{2}
\end{align}
for all $i=1,\cdots, n$. Note that since the maximum value of the triple CHSH inequality is $6\sqrt{2}$ and that the above is a convex mixture of triple CHSH inequalities for different $k,l$, for each $k,l$ we have
\begin{align}
&\bra{\psi}\bigg[\Z_i^{(k)}(\D^{(l)}_{\tx{zx},i}+\E^{(l)}_{\tx{zx},i})+\X^{(k)}_i(\D^{(l)}_{\tx{zx},i}-\E^{(l)}_{\tx{zx},i}) + \Z^{(k)}_i(\D^{(l)}_{\tx{zy},i}+\E^{(l)}_{\tx{zy},i}) \\ \nonumber & \quad\quad-\Y^{(k)}_i(\D^{(l)}_{\tx{zy},i}-\E^{(l)}_{\tx{zy},i}) +\X^{(k)}_i(\D^{(l)}_{\tx{xy},i}+\E^{(l)}_{\tx{xy},i})-\Y^{(k)}_i(\D^{(l)}_{\tx{xy},i}-\E^{(l)}_{\tx{xy},i})\bigg]\ket{\psi}=6\sqrt{2}.
\end{align}
Now, we may again use the SOS decomposition \eqref{chshsos} for each $i,k,l$ leading to 
\begin{align}\label{sosk1}
\Z_i^{(k)}\ket{\psi}&=\frac{\D^{(l)}_{\tx{zx},i}+\E^{(l)}_{\tx{zx},i}}{\sqrt{2}}\ket{\psi} =  \frac{\D^{(l)}_{\tx{zy},i}+\E^{(l)}_{\tx{zy},i}}{\sqrt{2}}\ket{\psi}, \\ 
\X_i^{(k)}\ket{\psi}&=\frac{\D^{(l)}_{\tx{zx},i}-\E^{(l)}_{\tx{zx},i}}{\sqrt{2}}\ket{\psi} =  \frac{\D^{(l)}_{\tx{xy},i}+\E^{(l)}_{\tx{xy},i}}{\sqrt{2}}\ket{\psi}, \\ 
\Y_i^{(k)}\ket{\psi}&=\frac{\D^{(l)}_{\tx{zy},i}-\E^{(l)}_{\tx{zy},i}}{\sqrt{2}}\ket{\psi} =  \frac{\D^{(l)}_{\tx{xy},i}-\E^{(l)}_{\tx{xy},i}}{\sqrt{2}}\ket{\psi}, \label{sosk2}
\end{align}
which we may write as  
\begin{align}\label{stcon1}
\Z_i^{(k)}\ket{\psi}=\Z_{i+n}^{(l)}\ket{\psi} ;\quad \X_i^{(k)}\ket{\psi}=\X_{i+n}^{(l)}\ket{\psi};\quad \Y_i^{(k)}\ket{\psi}&=\Y_{i+n}^{(l)}\ket{\psi},
\end{align}
where
\begin{align}
\Z_{i+n}^{(l)}=\frac{\D^{(l)}_{\tx{zx},i}+\E^{(l)}_{\tx{zx},i}}{\sqrt{2}}\, ,\quad \X_{i+n}^{(l)} = \frac{\D^{(l)}_{\tx{zx},i}-\E^{(l)}_{\tx{zx},i}}{\sqrt{2}}, \, \quad \Y_i^{(k)}\ket{\psi}&=\frac{\D^{(l)}_{\tx{zy},i}-\E^{(l)}_{\tx{zy},i}}{\sqrt{2}}.
\end{align}
As before, equations \eqref{sosk1} -- \eqref{sosk2} imply mutual anti-communtation of Alice's operators:
\begin{align}\label{comgen}
\{\Z_i^{(k)},\X_i^{(k)}\}=0 ;\quad \{\Z_i^{(k)},\Y_i^{(k)}\}=0 ; \quad \{\X_i^{(k)},\Y_i^{(k)}\} =0 \quad \forall i, k 
\end{align}
Defining 
\begin{align}
\Z_{i+n}=\frac{1}{6^{n-1}}\sum_l\Z_{i+n}^{(l)}; \quad\X_{i+n}=\frac{1}{6^{n-1}}\sum_l\X_{i+n}^{(l)}; \quad \Y_{i+n}=-\frac{1}{6^{n-1}}\sum_l\Y_{i+n}^{(l)} 
\end{align}
we have from \eqref{stcon1}
\begin{align}\label{zzstatement}
\Z_i^{(k)}\ket{\psi}=\Z_{i+n}\ket{\psi} ;\quad \X_i^{(k)}\ket{\psi}=\X_{i+n}\ket{\psi};\quad \Y_i^{(k)}\ket{\psi}&=-\Y_{i+n}\ket{\psi}
\end{align}
for all $k$. Note that the operators $\Z_{i+n}$, $\X_{i+n}$, $\Y_{i+n}$ are not necessarily unitary. We therefore define the regularized versions of these operators, denoted by $\hat{\Z}_{i+n}$, $\hat{\X}_{i+n}$ and $\hat{\Y}_{i+n}$, which using standard techniques (see for example \cite{BampsS,SASA}) can be shown to have the same properties: 
\begin{align}\label{stconzz}
\Z_i^{(k)}\ket{\psi}=\hat{\Z}_{i+n}\ket{\psi} ;\quad \X_i^{(k)}\ket{\psi}=\hat{\X}_{i+n}\ket{\psi};\quad \Y_i^{(k)}\ket{\psi}&=-\hat{\Y}_{i+n}\ket{\psi}. 
\end{align}
At this point we are nearly ready to construct our self-testing unitary. However, we still need to prove that $P^{(k)}_i$ and $P^{(k)}_j$ for $P\in\{\X,\Y,\Z\}$ commute for $i\neq j$. Here, we again use the method of \cite{ColS} to achieve this, which we restate here. Note that for every $i\neq j$, if we fix $z_i=1$ and $z_j=1$, there are $3^{n-2}$ choices for Charlie's measurement vector $\vz$. There are thus $3^{n-2}$ pairs of indices $(k,k')$ such that operators $\Z_i^{(k)}$ and $\Z_i^{(k')}$ are built from the same set of orthogonal projectors that commute by construction. We thus have $3^{n-2}$ equations of the form 
\begin{align}\label{kkcommute}
\Z_i^{(k)}\Z_j^{(k')}\ket{\psi}=\Z_j^{(k')}\Z_i^{(k)}\ket{\psi}.
\end{align}
Choosing a pair $(k,k')$ and using \eqref{zzstatement} and the fact that operators on Chalie and Alice's subsystems commute we then obtain
\begin{align}
\Z_i^{(k)}\Z_{n+j}\ket{\psi}&=\Z_j^{(k')}\Z_{n+i}\ket{\psi} \\
\Z_{n+j}\Z_i^{(k)}\ket{\psi}&=\Z_{n+i}\Z_j^{(k')}\ket{\psi}\\
\Z_{n+j}\Z_{n+i}\ket{\psi}&=\Z_{n+i}\Z_{n+j}\ket{\psi}.
\end{align}
In fact, by working backwards using different values of $k$, $k'$ and \eqref{zzstatement} again, one sees
\begin{align}\label{zzcom}
\Z_i^{(k)}\Z_j^{(k')}\ket{\psi}=\Z_j^{(k')}\Z_i^{(k)}\ket{\psi} \quad\quad \forall\, k,k',i\neq j.
\end{align}
In the same fashion, one proves 
\begin{align}
\X_i^{(k)}\X_j^{(k')}\ket{\psi}=\X_j^{(k')}\X_i^{(k)}\ket{\psi} \quad\quad \forall\, k,k',i\neq j, \label{xxcom}\\
\Y_i^{(k)}\Y_j^{(k')}\ket{\psi}=\Y_j^{(k')}\Y_i^{(k)}\ket{\psi} \quad\quad \forall\, k,k',i\neq j, \label{yycom}\\
\X_i^{(k)}\Y_j^{(k')}\ket{\psi}=\Y_j^{(k')}\X_i^{(k)}\ket{\psi} \quad\quad \forall\, k,k',i\neq j,
\label{xycom}\\
\X_i^{(k)}\Z_j^{(k')}\ket{\psi}=\Z_j^{(k')}\X_i^{(k)}\ket{\psi} \quad\quad \forall\, k,k',i\neq j,
\label{xzcom}\\
\Y_i^{(k)}\Z_j^{(k')}\ket{\psi}=\Z_j^{(k')}\Y_i^{(k)}\ket{\psi} \quad\quad \forall\, k,k',i\neq j,
\label{yzcom}
\end{align}
We have now finished the necessary groundwork to construct the self-testing circuit of Lemma 2.


\begin{figure*}
\centering
\includegraphics[scale=1]{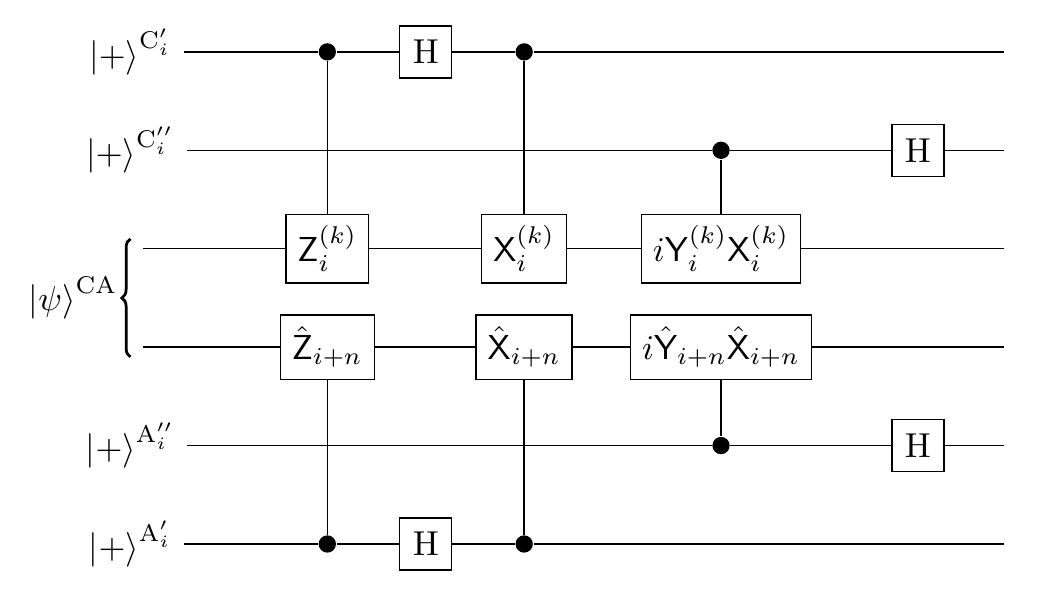}
\caption{\label{circuitgen} Circuit diagram representing the local unitary of Lemma 2. The total unitary consists of applying this circuit for each $i=1,\cdots,n$, and $k$ can be chosen to be any number $k=1,\cdots,3^{n-1}$ (for example $k=1$).}
\label{fig:swap}
\end{figure*}

\subsection{Verification of circuit}
The circuit we use (see Fig.\ \ref{circuitgen})  is a parallel version of the circuit used in the two qubit case. To prove that it functions correctly, we make repeated use of the properties \eqref{comgen}, \eqref{stconzz} and \eqref{zzcom} - \eqref{yzcom}.  Before the action of the first controlled gate the system is in state
\begin{equation}
\ket{\psi}^{\rC\rA}\frac{1}{2^{2n}}\sum_{p,q,r,s \in (0,1)^n}\ket{p}^{\rC'}\ket{q}^{\rC''}\ket{r}^{\rA'}\ket{s}^{\rA''},
\end{equation}
and after the first controlled gate the state evolves to
\begin{equation}
\frac{1}{2^{2n}}\sum_{p,q,r,s \in (0,1)^n}\left[\tp_{i=1}^n (\Z_i^{(k)})^{p_i}(\hat{\Z}_{i+n})^{r_i}\ket{\psi}^{\rC\rA}\right]\ket{p}^{\rC'}\ket{q}^{\rC''}\ket{r}^{\rA'}\ket{s}^{\rA''},
\end{equation}
where $p_i(r_i)$ is the $i$-th element of string $p(r)$. Hadamard gates evolve the state to
\begin{equation}
\frac{1}{2^{3n}}\sum_{p,q,r,s \in (0,1)^n}\left[\tp_{i=1}^n (\idd + (-1)^{p_i}\Z_i^{(k)})(\idd + (-1)^{r_i}\hat{\Z}_{i+n})\ket{\psi}^{\rC\rA}\right]\ket{p}^{\rC'}\ket{q}^{\rC''}\ket{r}^{\rA'}\ket{s}^{\rA''},
\end{equation}
and the second controlled gates lead to
\begin{equation}
\frac{1}{2^{3n}}\sum_{p,q,r,s \in (0,1)^n}\left[\tp_{i=1}^n (\X_i^{(k)})^{p_i}(\idd + (-1)^{p_i}\Z_i^{(k)})(\hat{\X}_{n+i})^{r_i}(\idd + (-1)^{r_i}\hat{\Z}_{i+n})\ket{\psi}^{\rC\rA}\right]\ket{p}^{\rC'}\ket{q}^{\rC''}\ket{r}^{\rA'}\ket{s}^{\rA''}.
\end{equation}
Relations (\ref{stconzz}) and \eqref{xzcom} allow us to simplify this to
\begin{equation}
\frac{1}{2^{3n}}\sum_{p,q,r,s \in (0,1)^n}\left[\tp_{i=1}^n (\X_i^{(k)})^{p_i}(\idd + (-1)^{p_i}\Z_i^{(k)})(\hat{\X}_{n+i})^{r_i}(\idd + (-1)^{r_i}\Z_{i}^{(k)})\ket{\psi}^{\rC\rA}\right]\ket{p}^{\rC'}\ket{q}^{\rC''}\ket{r}^{\rA'}\ket{s}^{\rA''}.
\end{equation}
Unitarity and hermiticity of $\Z_i^{(k)}$ implies $(\idd + \Z_i^{(k)})(\idd - \Z_i^{(k)})\ket{\psi} = 0$  and $\frac{1}{4}(\idd + \Z_i^{(k)})(\idd + \Z_i^{(k)})\ket{\psi} = \frac{1}{2}(\idd + \Z_i^{(k)})\ket{\psi}$ so that for every $i$ the state of the system can be further simplified to obtain
\begin{equation}
\frac{1}{2^{2n}}\sum_{p,q,s \in (0,1)^n}\left[\tp_{i=1}^n (\X_i^{(k)})^{p_i}(\idd + (-1)^{p_i}\Z_i^{(k)})(\hat{\X}_{n+i})^{p_i}\ket{\psi}^{\rC\rA}\right]\ket{p}^{\rC'}\ket{q}^{\rC''}\ket{p}^{\rA'}\ket{s}^{\rA''}.
\end{equation}
This can be further simplified by using \eqref{comgen} and \eqref{xxcom}:
\begin{multline}
\frac{1}{2^{2n}}\sum_{p,q,s \in (0,1)^n}\left[\tp_{i=1}^n (\idd + \Z_i^{(k)})\ket{\psi}^{\rC\rA}\right]\ket{p}^{\rC'}\ket{q}^{\rC''}\ket{p}^{\rA'}\ket{s}^{\rA''}  \\ \quad\quad = \frac{1}{2^{\frac{3n}{2}}}\sum_{q,s \in (0,1)^n}\left[\tp_{i=1}^n (\idd + \Z_i^{(k)})\ket{\psi}^{\rC\rA}\right]\left[\tp_{i=1}^n\ket{\Phi^{\tx{+}}}^{C_i'A_i'}\right]\ket{q}^{\rC''}\ket{s}^{\rA''}.
\end{multline}
Already here the state of the primed auxiliarys (extraction auxiliarys in the following text) is $n$-fold tensor product of maximally entangled pairs of qubits. Since the rest of the circuit does not affect extraction auxiliarys for the sake of simplicity it will be omitted from the following expressions. Following the action of the third pair of controlled gates the system evolves to
\begin{equation}
\frac{1}{2^{\frac{3n}{2}}}\sum_{q,s \in (0,1)^n}\left[\tp_{i=1}^n (i\Y_i^{(k)}\X_i^{(k)})^{q_i}(\idd + \Z_i^{(k)})(i\hat{\Y}_{n+i}\hat{\X}_{n+i})^{s_i}\ket{\psi}^{\rC\rA}\right]\ket{q}^{\rC''}\ket{s}^{\rA''},
\end{equation}
By virtue of \eqref{stconzz}, \eqref{comgen}, \eqref{yzcom}, \eqref{xycom} and \eqref{xzcom} this simplifies to
\begin{equation}
\frac{1}{2^{\frac{3n}{2}}}\sum_{q,s \in (0,1)^n}\left[\tp_{i=1}^n (i\Y_i^{(k)}\X_i^{(k)})^{q_i+s_i}(\idd + \Z_i^{(k)})\ket{\psi}^{\rC\rA}\right]\ket{q}^{\rC''}\ket{s}^{\rA''},
\end{equation}
Finally at the end of the circuit, after the action of the second pair of Hadamards we have:
\begin{equation}
\frac{1}{2^{\frac{5n}{2}}}\sum_{q,s,\bar{q},\bar{s} \in (0,1)^n}\left[\tp_{i=1}^n(-1)^{\bar{q}_iq_i + \bar{s}_is_i} (i\Y_i^{(k)}\X_i^{(k)})^{q_i+s_i}(\idd + \Z_i^{(k)})\ket{\psi}^{\rC\rA}\right]\ket{\bar{q}}^{\rC''}\ket{\bar{s}}^{\rA''}.
\end{equation}
Note that each term from the sum is characterised by a pair of strings $(\bar{q},\bar{s})$ and a set of pairs of strings $\Xi$, such that  $q''_j + s''_j = q'_j + s'_j$ for every $q'',s'',q',s' \in \Xi$ and every $j$. We show that the multiplicative factor in front of every term is equal to zero whenever $\bar{q}'\neq \bar{s}'$. Let us assume $\bar{q}' = \bar{s}'$. The multiplicative factor for a term corresponding to a pair of strings $q',s'$ is equal to
\begin{equation*}
(-1)^{\sum_{q',s'\in \Xi, j}\bar{q}'_jq'_j + \bar{s}'_js'_j} = (-1)^{\sum_{q',s'\in \Xi, j}\bar{q}'_j(q'_j+s'_j)} = \pm 1,
\end{equation*}
i.e., all the terms come with the same sign, since sum is over $q',s'$ which have fixed $q'_j + s'_j$ for every $j$. 
Contrarily, in case $\bar{q}' \neq \bar{s}'$ the multiplicative factor for a term corresponding to a pair of strings $q',s'$ is equal to
\begin{align*}
(-1)^{\sum_{q',s'\in \Xi, j}\bar{q}'_jq'_j + \bar{s}'_js'_j} &= (-1)^{\sum_{q',s'\in \Xi, j}\bar{q}'_j(q'_j+s'_j) + (\bar{s}_j'-\bar{q}'_j)s'_j} = \begin{cases} \pm 1 \quad \textrm{when} \quad  \sum_js'_j = 0 \\ \mp 1   \quad \textrm{when} \quad  \sum_js'_j = 1 \end{cases}\\
&= 0.
\end{align*}
In this case value of $s'_j$ determines the sign of the terms, and for half of the terms it is equal $0$ (one sign) and for the half it is equal to $1$ (opposite sign).  This means that only terms of the sum which survive are those corresponding to $\bar{q} = \bar{s}$.
\begin{equation}
\frac{1}{2^{\frac{5n}{2}}}\sum_{q,s,\bar{q} \in (0,1)^n}\left[\tp_{i=1}^n(-1)^{\bar{q}_i(q_i + s_i)} (i\Y_i^{(k)}\X_i^{(k)})^{q_i+s_i}(\idd + \Z_i^{(k)})\ket{\psi}^{\rC\rA}\right]\ket{\bar{q}\bar{q}}^{\rC''\rA''}.
\end{equation}
The sum has $2^{3n}$ different contributions (one for each triple $q,s,\bar{q}$), but there are $2^{2n}$ different terms, meaning that each term has contributions from $2^{n}$ different pairs of strings $(q,s)$. This reduces the multiplicative factor in front of the sum to ${2^{-\frac{3n}{2}}}$. After summing over $q,s$  and making some rearrangements the expression reduces to
%
%
%
%
%
\begin{equation}\label{xi}
\ket{\xi} = \frac{1}{2^{\frac{3n}{2}}}\sum_{\bar{q} \in (0,1)^n}\left[\tp_{i=1}^n (\idd + (-1)^{\bar{q}_i}i\Y_i^{(k)}\X_i^{(k)})(\idd + \Z_{i}^{(k)})\ket{\psi}^{\rC\rA}\right]\ket{\bar{q}\bar{q}}^{\rC''\rA''}.
\end{equation}
Finally, by returning the state of extraction auxiliary systems one obtains the statement from Lemma 2:
\begin{equation}
U\left[\ket{\psi}^{\rC\rA}\otimes\ket{00}\right] = \ket{\xi} \tp_{i=1}^n {\ket{\Phi^{\tx{+}}}}^{\rC_i'\rA_i'}. \label{2}
\end{equation}
Before calculating the output of the circuit when the input is $\Z_{i}^{(k)}\ket{\psi}$ let us acknowledge that $\Z_{i}^{(k)}\ket{\psi} = \Z_{i}^{(l)}\ket{\psi}$ for any two $l$ and $k$, which can be seen from \eqref{stconzz} which is satisfied for any $k$. The same holds for $\X_{i}^{(k)}\ket{\psi}$ and $\Y_{i}^{(k)}\ket{\psi}$. By repeating the same procedure as in the derivation above one can confirm two more statements from Lemma 2 for any $k$ and $j$:
\begin{eqnarray}
U\left[\Z_j^{(k)}\ket{\psi}^{\rC\rA}\otimes\ket{00}\right]&=& \ket{\xi} \left[ {\sigma_z}^{\rC_j'} \tp_{i=1}^n{\ket{\Phi^{\tx{+}}}}^{\rC_i'\rA_i'}\right],\\ \nonumber
U\left[\X_j^{(k)}\ket{\psi}^{\rC\rA}\otimes\ket{00}\right]&=& \ket{\xi}\left[{\sigma_x}^{\rC_j'}\tp_{i=1}^n{\ket{\Phi^{\tx{+}}}}^{\rC_i'\rA_i'}\right].\\ \nonumber
\end{eqnarray}
The situation when the input state is $\Y_j^{(k)}\ket{\psi}$ is a bit more complicated so more details of the derivation will be presented. After the second pair of controlled gates the state of the system is:
\begin{equation}
\frac{1}{2^{3n}}\sum_{p,q,r,s \in (0,1)^n}\left[\tp_{i=1}^n (\X_i^{(k)})^{p_i}(\idd + (-1)^{p_i}\Z_i^{(k)})\Y_j^{(k)}(\hat{\X}_{n+i})^{r_i}(\idd + (-1)^{r_i}\hat{\Z}_{i+n})\ket{\psi}^{\rC\rA}\right]\ket{p}^{\rC'}\ket{q}^{\rC''}\ket{r}^{\rA'}\ket{s}^{\rA''},
\end{equation}
which due to eqs. (\ref{comgen}) and (\ref{yzcom}) simplifies to
\begin{equation}
\frac{1}{2^{3n}}\sum_{p,q,r,s \in (0,1)^n}\left[\tp_{i=1}^n (\X_i^{(k)})^{p_i}\Y_j^{(k)}(\idd + (-1)^{p_i \oplus \delta_{ij}}\Z_i^{(k)})(\hat{\X}_{n+i})^{r_i}(\idd + (-1)^{r_i}\hat{\Z}_{i+n})\ket{\psi}^{\rC\rA}\right]\ket{p}^{\rC'}\ket{q}^{\rC''}\ket{r}^{\rA'}\ket{s}^{\rA''},
\end{equation}
By using \eqref{zzcom}, \eqref{comgen} and the fact that $\frac{\idd + \Z_i^{(k)}}{2}$ and $\frac{\idd - \Z_i^{(k)}}{2}$ are projectors onto different eigenspaces of $\Z_i^{(k)}$ the above reduces to
\begin{equation}
\frac{1}{2^{2n}}\sum_{q,r,s \in (0,1)^n}\left[\tp_{i=1}^n (-1)^{r_i\oplus \delta_{ij}}\Y_j^{(k)}\X_{j}^{(k)}(\idd + \Z_i^{(k)})\ket{\psi}^{\rC\rA}\right]\ket{r\oplus 1_j}^{\rC'}\ket{q}^{\rC''}\ket{r}^{\rA'}\ket{s}^{\rA''},
\end{equation}
where $1_j$ is an $n$-element string whose $j$-th element is one with all the other elements being zeros. The last expression can be rewritten in the following way:
\begin{equation}
\frac{1}{2^{\frac{3n}{2}}}\sum_{q,s \in (0,1)^n}\left[\tp_{i=1}^n i\Y_j^{(k)}\X_{j}^{(k)}(\idd + \Z_i^{(k)})\ket{\psi}^{\rC\rA}\right]\sigma_y^{C_j'}\left[\tp_{i=1}^n\ket{\Phi^{\tx{+}}}^{C_i'A_i'}\right]\ket{q}^{\rC''}
\ket{s}^{\rA''}.
\end{equation}
Since the rest of the circuit does not affect the state of extraction auxiliarys we will drop it from the following few equations. After applying the third pair of controlled gates on this state one obtains
\begin{equation}
\frac{1}{2^{\frac{3n}{2}}}\sum_{q,s \in (0,1)^n}\left[\tp_{i=1}^n (i\Y_i^{(k)}\X_{i}^{(k)})^{q_i + \delta_{ij}}(\idd + \Z_i^{(k)})(i\hat{\Y}_{i+n}\hat{\X}_{i+n})^{s_i}\ket{\psi}^{\rC\rA}\right]\ket{q}^{\rC''}\ket{s}^{\rA''},
\end{equation}
which due to \eqref{stconzz} and anticommuting relations \eqref{comgen} reduces to:
\begin{equation}
\frac{1}{2^{\frac{3n}{2}}}\sum_{q,s \in (0,1)^n}\left[\tp_{i=1}^n (i\Y_i^{(k)}\X_{i}^{(k)})^{s_i + q_i + \delta_{ij}}(\idd + \Z_i^{(k)})\ket{\psi}^{\rC\rA}\right]\ket{q}^{\rC''}\ket{s}^{\rA''},
\end{equation}
and at the end of the circuit following the action of two last Hadamards this state transforms to
\begin{equation}
\frac{1}{2^{\frac{5n}{2}}}\sum_{\bar{q},\bar{s},q,s \in (0,1)^n}(-1)^{\bar{q}_iq_i+\bar{s}_is_i}\left[\tp_{i=1}^n (i\Y_i^{(k)}\X_{i}^{(k)})^{q_i+s_i + \delta_{ij}}(\idd + \Z_i^{(k)})\ket{\psi}^{\rC\rA}\right]\ket{\bar{q}}^{\rC''}\ket{s}^{\rA''}.
\end{equation}
Here the same reasoning like the one preceding to eq. (\ref{xi}) can be applied, the only difference being factor $(i\Y_i^{(k)}\X_{i}^{(k)})^{\delta_{ij}}$. This factor changes the sign of terms in (\ref{xi}) which correspond to any string $\bar{q}$ for which $\bar{q}_j = 1$. The final form of the output of the circuit when input is $\Y_j^{(k)}\ket{\psi}$ can be written as 
\begin{equation}\label{xiY}
\frac{1}{2^{\frac{3n}{2}}}\sum_{\bar{q} \in (0,1)^n}\left[\tp_{i=1}^n (-1)^{\bar{q}_j}(\idd + (-1)^{\bar{q}_i}i\Y_i^{(k)}\X_i^{(k)})(\idd + \Z_{i}^{(k)})\ket{\psi}^{\rC\rA}\right]\sigma_y^{\rC_j'}\left[\tp_{i=1}^n\ket{\Phi^{\tx{+}}}^{\rC_i'\rA_i'}\right]\ket{\bar{q}\bar{q}}^{\rC''\rA''},
\end{equation}
which is equivalent to the formulation from Lemma 2:
\begin{equation}
U\left[\Y_j^\rC\ket{\psi}^{\rC\rA}\otimes\ket{00}\right]= {\sigma_z}^{\rC_j''}\ket{\xi}\left[{\sigma_y}^{\rC_j'}\tp_{i=1}^n{\ket{\Phi^{\tx{+}}}}^{\rC_i'\rA_i'}\right]
\end{equation}
which completes the proof.


\section{Proof of Lemma \ref{BSMlemma}}\label{bsmproof}
Correlations $\bra{\psi}\Ss_{l,a}\ket{\psi} = \bra{\psi}\Ss_{l,a}\ket{\psi} = \frac{1}{4}$ for every $l \in \{1,\dots m\}$ and $a\in \{0,1,2,3\}$, given in Table \ref{bsmtab1}, imply that the norm of states $\Ss_{l,a}\ket{\psi}$ and $\T_{l,a}\ket{\psi}$ is equal to $\frac{1}{2}$. These correlations allow us to write
\begin{equation}\label{sim}
\Ss_{l,0}\ket{\psi} \sim \frac{1}{4}\left(\ket{\psi} +  \Z_{2l-1}^{(k)}\Z_{2l}^{(k)}\ket{\psi} + \X_{2l-1}^{(k)}\X_{2l}^{(k)}\ket{\psi} - \Y_{2l-1}^{(k)}\Y_{2l}^{(k)}\ket{\psi}\right).
\end{equation}
Since states $\ket{\psi}$, $\Z_{2l-1}^{(k)}\Z_{2l}^{(k)}\ket{\psi}$, $\X_{2l-1}^{(k)}\X_{2l}^{(k)}\ket{\psi}$ and $\Y_{2l-1}^{(k)}\Y_{2l}^{(k)}\ket{\psi}$ all have unit norm and are mutually orthogonal they can be seen as a part of basis of all states from $\mathcal{H}^C\tp \mathcal{H}^A$. Moreover $\Ss_{l,0}\ket{\psi}$  has the same norm as the expression from the right hand side of $\sim$  in eq. (\ref{sim})  which implies that
\begin{equation}\label{M0}
\Ss_{l,0}\ket{\psi} = \frac{1}{4}\left(\ket{\psi} +  \Z_{2l-1}^{(k)}\Z_{2l}^{(k)}\ket{\psi} + \X_{2l-1}^{(k)}\X_{2l}^{(k)}\ket{\psi} - \Y_{2l-1}^{(k)}\Y_{2l}^{(k)}\ket{\psi}\right).
\end{equation}
The same reasoning leads to the following set of equations:
\begin{eqnarray}\label{M1}
\Ss_{l,1}\ket{\psi} &=& \frac{1}{4}\left(\ket{\psi} +  \Z_{2l-1}^{(k)}\Z_{2l}^{(k)}\ket{\psi} - \X_{2l-1}^{(k)}\X_{2l}^{(k)}\ket{\psi} + \Y_{2l-1}^{(k)}\Y_{2l}^{(k)}\ket{\psi}\right),\\ \label{M2}
\Ss_{l,2}\ket{\psi} &=& \frac{1}{4}\left(\ket{\psi} -  \Z_{2l-1}^{(k)}\Z_{2l}^{(k)}\ket{\psi} + \X_{2l-1}^{(k)}\X_{2l}^{(k)}\ket{\psi} + \Y_{2l-1}^{(k)}\Y_{2l}^{(k)}\ket{\psi}\right),\\ \label{M3}
\Ss_{l,3}\ket{\psi} &=& \frac{1}{4}\left(\ket{\psi} -  \Z_{2l-1}^{(k)}\Z_{2l}^{(k)}\ket{\psi} - \X_{2l-1}^{(k)}\X_{2l}^{(k)}\ket{\psi} - \Y_{2l-1}^{(k)}\Y_{2l}^{(k)}\ket{\psi}\right),\\ \label{N0}
\T_{l,0}\ket{\psi} &=& \frac{1}{4}\left(\ket{\psi} +  \Z_{2l}^{(k)}\Z_{2l+1}^{(k)}\ket{\psi} + \X_{2l}^{(k)}\X_{2l+1}^{(k)}\ket{\psi} - \Y_{2l}^{(k)}\Y_{2l+1}^{(k)}\ket{\psi}\right),\\  \label{N1}
\T_{l,1}\ket{\psi} &=& \frac{1}{4}\left(\ket{\psi} +  \Z_{2l}^{(k)}\Z_{2l+1}^{(k)}\ket{\psi} - \X_{2l}^{(k)}\X_{2l+1}^{(k)}\ket{\psi} + \Y_{2l}^{(k)}\Y_{2l+1}^{(k)}\ket{\psi}\right),\\  \label{N2}
\T_{l,2}\ket{\psi} &=& \frac{1}{4}\left(\ket{\psi} -  \Z_{2l}^{(k)}\Z_{2l+1}^{(k)}\ket{\psi} + \X_{2l}^{(k)}\X_{2l+1}^{(k)}\ket{\psi} + \Y_{2l}^{(k)}\Y_{2l+1}^{(k)}\ket{\psi}\right),\\  \label{N3}
\T_{l,3}\ket{\psi} &=& \frac{1}{4}\left(\ket{\psi} -  \Z_{2l}^{(k)}\Z_{2l+1}^{(k)}\ket{\psi} - \X_{2l}^{(k)}\X_{2l+1}^{(k)}\ket{\psi} - \Y_{2l}^{(k)}\Y_{2l+1}^{(k)}\ket{\psi}\right).
\end{eqnarray}
Equations (\ref{M0}-\ref{M3}) are equivalent to the following set of equations
\begin{subequations}
\begin{eqnarray}\label{Zz}
\Z_{2l-1}^{(k)}\Z_{2l}^{(k)}\ket{\psi} &=& \left(\Ss_{l,0} + \Ss_{l,1} - \Ss_{l,2} - \Ss_{l,3}\right)\ket{\psi}, \\ \label{Xx}
\X_{2l-1}^{(k)}\X_{2l}^{(k)}\ket{\psi} &=& \left(\Ss_{l,0} - \Ss_{l,1} + \Ss_{l,2} - \Ss_{l,3}\right)\ket{\psi}, \\ \label{Yy}
\Y_{2l-1}^{(k)}\Y_{2l}^{(k)}\ket{\psi} &=& \left(-\Ss_{l,0} + \Ss_{l,1} + \Ss_{l,2} - \Ss_{l,3}\right)\ket{\psi}.
\end{eqnarray}
\end{subequations}
Based on the last set of equations and the fact that $\{\Ss_{l,a}\}_{l,a}$ is orthogonal set of projectors which all commute with all the operators from $\{\Z_{j}^{(k)},\X_{j}^{(k)}\}_{j,k}$ one can show that
\begin{eqnarray}\nonumber
\X_{2l-1}^{(k)}\X_{2l}^{(k)}\Z_{2l-1}^{(k)}\Z_{2l}^{(k)}\ket{\psi} &=& \X_{2l-1}^{(k)}\X_{2l}^{(k)}\left(\Ss_{l,0} + \Ss_{l,1} - \Ss_{l,2} - \Ss_{l,3}\right)\ket{\psi}\\ \nonumber
&=& \left(\Ss_{l,0} + \Ss_{l,1} - \Ss_{l,2} - \Ss_{l,3}\right)\left(\Ss_{l,0} - \Ss_{l,1} + \Ss_{l,2} - \Ss_{l,3}\right)\ket{\psi}\\ \nonumber
&=& \left(\Ss_{l,0}- \Ss_{l,1} - \Ss_{l,2} + \Ss_{l,3} \right)\ket{\psi}\\ \label{XZY}
&=& -\Y_{2l-1}^{(k)}\Y_{2l}^{(k)}\ket{\psi}.
\end{eqnarray}
Starting from equations (\ref{N0}-\ref{N3}) one can obtain
\begin{equation}\label{XZY+1}
\X_{2l}^{(k)}\X_{2l+1}^{(k)}\Z_{2l}^{(k)}\Z_{2l+1}^{(k)}\ket{\psi} = -\Y_{2l}^{(k)}\Y_{2l+1}^{(k)}\ket{\psi}.
\end{equation}
Equations (\ref{XZY}) and (\ref{XZY+1}) hold for every $k$ and every $l$. Let us take $l=1$ and check how eq. (\ref{XZY}) affects vector $\ket{\xi_{\bar{q}}} = \tp_{i=1}^n(\idd + (-1)^{\bar{q}_i}i\Y_i^{(k)}\X_i^{(k)})(\idd + \Z_{i}^{(k)})\ket{\psi}$. Let us write it in the following form:
\begin{equation*}
\ket{\xi_{\bar{q}}} = L_{\textrm{rest}}\tp\left(\idd + (-1)^{\bar{q}_1}i\Y_1^{(k)}\X_1^{(k)}\right)\left(\idd + \Z_{1}^{(k)}\right)\tp\left(\idd + (-1)^{\bar{q}_2}i\Y_2^{(k)}\X_2^{(k)}\right)\left(\idd + \Z_{2}^{(k)}\right) \ket{\psi}
\end{equation*}
where $L_{\textrm{rest}} = \tp_{i=3}^n (\idd + (-1)^{\bar{q}_i}i\Y_i^{(k)}\X_i^{(k)})(\idd + \Z_{i}^{(k)})$. Let us assume $\bar{q}_1 \neq \bar{q}_2$ and omit $L_{\textrm{rest}}$ for the sake of shorter exposition. Then $\ket{\xi_{\bar{q}}}$ reads
\begin{eqnarray*}
\ket{\psi} \pm i\Y_2^{(k)}\X_2^{(k)}\ket{\psi} &+& \Z_{2}^{(k)}\ket{\psi} \pm i\Y_2^{(k)}\X_2^{(k)}\Z_{2}^{(k)}\ket{\psi}
\mp i\Y_1^{(k)}\X_1^{(k)}\ket{\psi} + \Y_1^{(k)}\X_1^{(k)}\Y_2^{(k)}\X_2^{(k)}\ket{\psi} \mp i\Y_1^{(k)}\X_1^{(k)}\Z_{2}^{(k)}\ket{\psi} + \\ + \Y_1^{(k)}\X_1^{(k)}\Y_2^{(k)}\X_2^{(k)}\Z_{2}^{(k)}\ket{\psi} &+& 
\Z_{1}^{(k)}\ket{\psi} \pm i\Z_{1}^{(k)}\Y_2^{(k)}\X_2^{(k)}\ket{\psi} + \Z_{1}^{(k)}\Z_{2}^{(k)}\ket{\psi} \pm  i\Z_{1}^{(k)}\Y_2^{(k)}\X_2^{(k)}\Z_{2}^{(k)}\ket{\psi} +
\mp i\Y_1^{(k)}\X_1^{(k)}\Z_{1}^{(k)}\ket{\psi} + \\ + \Y_1^{(k)}\X_1^{(k)}\Z_{1}^{(k)}\Y_2^{(k)}\X_2^{(k)}\ket{\psi} &+& \mp i\Y_1^{(k)}\X_1^{(k)}\Z_{1}^{(k)}\Z_{2}^{(k)}\ket{\psi} + \Y_1^{(k)}\X_1^{(k)}\Z_{1}^{(k)}\Y_2^{(k)}\X_2^{(k)}\Z_{2}^{(k)}\ket{\psi}.
\end{eqnarray*}
This expression can be written as a sum of expressions, each equal to $0$. To show this let us rearrange eq. (\ref{XZY}) for the case $l=1$. It can be written in eight different ways, which are given below.
\begin{align}\label{lancel}
\begin{split}
\ket{\psi} + \Y_1^{(k)}\X_1^{(k)}\Z_{1}^{(k)}\Y_2^{(k)}\X_2^{(k)}\Z_{2}^{(k)}\ket{\psi} = 0, &\qquad  \Y_2^{(k)}\X_2^{(k)}\ket{\psi} + \Y_1^{(k)}\X_1^{(k)}\Z_{1}^{(k)}\Z_{2}^{(k)}\ket{\psi} = 0, \\
\Z_{2}^{(k)}\ket{\psi} + \Y_1^{(k)}\X_1^{(k)}\Z_{1}^{(k)}\Y_2^{(k)}\X_2^{(k)}\ket{\psi} = 0, &\qquad  \Y_2^{(k)}\X_2^{(k)}\Z_{2}^{(k)}\ket{\psi} + \Y_1^{(k)}\X_1^{(k)}\Z_{1}^{(k)}\ket{\psi} = 0, \\ 
\Y_1^{(k)}\X_1^{(k)}\ket{\psi} + \Z_{1}^{(k)}\Y_2^{(k)}\X_2^{(k)}\Z_{2}^{(k)}\ket{\psi} = 0, &\qquad  \Y_1^{(k)}\X_1^{(k)}\Y_2^{(k)}\X_2^{(k)}\ket{\psi} + \Z_{1}^{(k)}\Z_{2}^{(k)}\ket{\psi} = 0, \\
\Y_1^{(k)}\X_1^{(k)}\Z_{2}^{(k)}\ket{\psi} + \Z_{1}^{(k)}\Y_2^{(k)}\X_2^{(k)}\ket{\psi} = 0,&\qquad  \Y_1^{(k)}\X_1^{(k)}\Y_2^{(k)}\X_2^{(k)}\Z_{2}^{(k)}\ket{\psi} + \Z_{1}^{(k)}\ket{\psi} = 0.
\end{split}
\end{align}
All these equations are obtained from eq. (\ref{XZY}) by using commutation relations (\ref{zzcom}), expressions \eqref{stconzz}, anti-commutation relations \eqref{comgen} and the fact that operators $P_{i}^{(k)}$ for $P \in \{X,Y,Z\}$ are reflections, defined by property ${P_{i}^{(k)}}^2 = \idd$ on the support of $\ket{\psi}$.\\ 

Premise $\bar{q}_1 \neq \bar{q}_2$ leads to conclusion $\ket{\xi_{\bar{q}}} = 0$. In a completely analogous way, starting from eq. (\ref{XZY}) one can show that $\ket{\xi_{\bar{q}}} = 0$ if there exists $l$ such that $\bar{q}_{2l-1} \neq \bar{q}_{2l}$.  Similarly, eq. (\ref{XZY+1}) can be used to prove that $\ket{\xi_{\bar{q}}} = 0$ if there exists $l$ such that $\bar{q}_{2l} \neq \bar{q}_{2l+1}$. The only two states $\ket{\bar{q}}$ which satisfy $\bar{q}_{2l-1} = \bar{q}_{2l} = \bar{q}_{2l+1}$ are $\ket{\bar{q}} = \ket{0\dots 0}$ and $\ket{\bar{q}} = \ket{1\dots 1}$. This means that
\begin{equation}\label{last}
\ket{\xi} = \ket{\xi_{0}}\tp\ket{0\dots 0} + \ket{\xi_{1}}\tp\ket{1\dots 1},
\end{equation}
which is exactly what had to be proven.


\section{Robust version of Lemma 3}\label{lemma3Rob}
In this section we show how one can derive a noise robust version of Lemma \ref{BSMlemma} given a noise robust version of Lemma \ref{lemma2}. Specifically, we show that if each of the probabilities differ by at most $\eta$ from the values in Table \ref{tablecors} one has
\begin{align} \label{bsmrob1}
\| U\left[\ket{\psi}\otimes\ket{00}\right] - \ket{\tilde\xi} \otimes \left[ \tp_{i=1}^n {\ket{\Phi^{\tx{+}}}}\right]\|\leq O(\epsilon^m)+O(\sqrt\eta),\\ 
\|U\left[\Pi_j \Z_j^\rC\ket{\psi}\otimes\ket{00}\right]- \ket{\tilde\xi} \otimes\left[ \otimes_j{\sigma_{\tx{z}}}^{\rC_j'} \tp_{i=1}^n{\ket{\Phi^{\tx{+}}}}\right]\|\leq O(\epsilon^m)+O(\sqrt\eta),\\ 
\|U\left[\Pi_j \X_j^\rC\ket{\psi}\otimes\ket{00}\right]-\ket{\tilde\xi}\otimes\left[\otimes_j{\sigma_{\tx{x}}}^{\rC_j'}\tp_{i=1}^n{\ket{\Phi^{\tx{+}}}}\right]\|\leq O(\epsilon^m)+O(\sqrt\eta),\\ 
\|U\left[\Pi_j \Y_j^\rC\ket{\psi}\otimes\ket{00}\right]- {\sigma_{\tx{z}}}^{\rC_j''}\ket{\tilde\xi}\otimes\left[\otimes_j{\sigma_{\tx{y}}}^{\rC_j'}\tp_{i=1}^n{\ket{\Phi^{\tx{+}}}}\right]\|\leq O(\epsilon^m)+O(\sqrt\eta),
\end{align}
where $\ket{\tilde\xi}$ is the state given in \eqref{newxi}
\begin{align}
\ket{\tilde\xi}=\ket{\xi_{0}}\tp\ket{0\dots 0} + \ket{\xi_{1}}\tp\ket{1\dots 1}
\end{align} 
and the scaling $\epsilon^m$ (for some $m$) follows from a robust self-test of Lemma \ref{lemma2} (see following) given non-maximal violation of the triple CHSH Bell inequalities. Here, we focus on proving \eqref{bsmrob1}; a similar technique can be applied to the remaining three equations. First, note that by writing ${\ket{\Phi^{\tx{+}}_n}}=\tp_{i=1}^n{\ket{\Phi^{\tx{+}}}}$ and using the triangle inequality we have 
\begin{align}
\| U\left[\ket{\psi}\otimes\ket{00}\right] - \ket{\tilde\xi} \otimes {\ket{\Phi^{\tx{+}}_n}}\|&= \| U\left[\ket{\psi}\otimes\ket{00}\right] - \ket{\xi} \otimes {\ket{\Phi^{\tx{+}}_n}} + \ket{\xi} \otimes {\ket{\Phi^{\tx{+}}_n}}-\ket{\xi} \otimes {\ket{\Phi^{\tx{+}}_n}} \|  \\ 
&\leq \| U\left[\ket{\psi}\otimes\ket{00}\right] - \ket{\tilde\xi} \otimes {\ket{\Phi^{\tx{+}}_n}}\| + \|\ket{\tilde\xi} \otimes {\ket{\Phi^{\tx{+}}_n}}  - \ket{\xi} \otimes {\ket{\Phi^{\tx{+}}_n}}\|,
\end{align}
where $\ket{\xi}$ is taken to be the state appearing in Lemma 2. The first term now gives the bound of order $\epsilon^m$ that follows from the robust self-test of Lemma \ref{lemma2}. We now focus on the second term, that is, we need to bound
\begin{equation*}
\|\ket{\xi} - (\ket{\xi_{0}}\tp\ket{0\dots 0} + \ket{\xi_{1}}\tp\ket{1\dots 1})\| \sim  O(\sqrt{\eta}).
\end{equation*}
%
Given that there is a positive $\eta$ such that observed probabilities are at most $\eta$ far from the values given in Table \ref{tablecors}, let us upper bound the following expression
\begin{align}\label{glavna}
\bigg\| \Ss_{l,0}\ket{\psi} - \frac{\mathbb{I} + \Z_{2l-1}\Z_{2l} + \X_{2l-1}\X_{2l}- \Y_{2l-1}\Y_{2l}}{4}\ket{\psi} \bigg\|.
\end{align}
By definition it is equal to 
\begin{align}\label{papir}
\begin{split}
\bigg(\langle \Ss_{l,0}\rangle - &\frac{\langle \Ss_{l,0}\rangle + \langle \Ss_{l,0}\Z_{2l-1}\Z_{2l}\rangle + \langle \Ss_{l,0}\X_{2l-1}\X_{2l}\rangle - \langle \Ss_{l,0}\Y_{2l-1}\Y_{2l}\rangle }{2} + \\ &+ \frac{\langle \mathbb{I}\rangle}{4} + \frac{ \langle\Z_{2l-1}\Z_{2l}\rangle + \langle\X_{2l-1}\X_{2l}\rangle- \langle\Y_{2l-1}\Y_{2l}\rangle}{8} + \\ &+
\frac{\langle\Z_{2l-1}\Z_{2l}\X_{2l-1}\X_{2l}\rangle + \langle\X_{2l-1}\X_{2l}\Z_{2l-1}\Z_{2l}\rangle - \langle\Z_{2l-1}\Z_{2l}\Y_{2l-1}\Y_{2l}\rangle}{16} + \\ &+ \frac{-\langle\Y_{2l-1}\Y_{2l}\Z_{2l-1}\Z_{2l}\rangle - \langle\X_{2l-1}\X_{2l}\Y_{2l-1}\Y_{2l}\rangle - \langle\Y_{2l-1}\Y_{2l}\X_{2l-1}\X_{2l}\rangle}{16}
 \bigg)^{\frac{1}{2}}
 \end{split}
\end{align}
Observe now that
\begin{align}\label{kamen}\begin{split}
& \vert \left(\bra{\psi}\otimes \bra{00}\right) \Z_{2l-1} \Z_{2l}^\rC\ket{\psi}\otimes\ket{00} \vert = \left| \left(\bra{\psi}\otimes \bra{00}\right)U^{\dagger} U\left(\Z_{2l-1} \Z_{2l}^\rC\ket{\psi}\otimes\ket{00}\right)\right| \\
&= \left| \left(\bra{\psi}\otimes \bra{00}\right)U^{\dagger}\left[U\left(\Z_{2l-1} \Z_{2l}^\rC\ket{\psi}\otimes\ket{00}\right) - \ket{\tilde\xi} \otimes\left[ {\sigma_{\tx{z}}}^{\rC_{2l-1}'}\tp{\sigma_{\tx{z}}}^{\rC_{2l}'}  \tp_{i=1}^n{\ket{\Phi^{\tx{+}}}}\right]  + \ket{\tilde\xi} \otimes\left[ {\sigma_{\tx{z}}}^{\rC_{2l-1}'}\tp{\sigma_{\tx{z}}}^{\rC_{2l}'}  \tp_{i=1}^n{\ket{\Phi^{\tx{+}}}}\right]\right] \right| \\
& \leq \left\|U\left[\ket{\psi}\otimes\ket{00}\right]\right\|\left\|U\left(\Z_{2l-1} \Z_{2l}^\rC\ket{\psi}\otimes\ket{00}\right) - \ket{\tilde\xi} \otimes\left[ {\sigma_{\tx{z}}}^{\rC_{2l-1}'}\tp{\sigma_{\tx{z}}}^{\rC_{2l}'}  \tp_{i=1}^n{\ket{\Phi^{\tx{+}}}}\right]  \right\| + \\ &\qquad + \left| \left( \left(\bra{\psi}\otimes \bra{00}\right)U^{\dagger} - \bra{\tilde\xi} \otimes \left[ \tp_{i=1}^n {\bra{\Phi^{\tx{+}}}}\right] +  \bra{\tilde\xi} \otimes \left[ \tp_{i=1}^n {\bra{\Phi^{\tx{+}}}}\right]\right)\ket{\tilde\xi} \otimes\left[ {\sigma_{\tx{z}}}^{\rC_{2l-1}'}\tp{\sigma_{\tx{z}}}^{\rC_{2l}'}  \tp_{i=1}^n{\ket{\Phi^{\tx{+}}}}\right] \right| \\
&\leq O(\epsilon^m) + \left\|U\left[\ket{\psi}\otimes\ket{00}\right] - \ket{\tilde\xi} \otimes \left[ \tp_{i=1}^n {\ket{\Phi^{\tx{+}}}}\right] \right\|\left\|\ket{\tilde\xi} \otimes\left[ \otimes_j{\sigma_{\tx{z}}}^{\rC_j'} \tp_{i=1}^n{\ket{\Phi^{\tx{+}}}}\right] \right\| \leq O(\epsilon^m).
\end{split}
\end{align}
In the first line we just added a unitary which does not change the inner product, while in the second line we just added a zero term. In the third line we used triangle and Cauchy-Schwartz inequalities. In the fourth line we again added a zero term and used again triangle and Cauchy-Schwartz inequalities to obtain the fifth line. Using the same sequence of steps the equivalent bound can be obtained for inner products of $ \langle \X_{2l-1} \X_{2l}\rangle$ and $ \langle \Y_{2l-1} \Y_{2l}\rangle$  and also for all inner products from the third and fourth line of \eqref{papir}. All these inner products have absolute value as the one derived in \eqref{kamen}.  Finally, to bound the first line from \eqref{papir} let us assume the worst case correction of Table 1, i.e
\begin{align*}
\langle \Ss_{0,l}\rangle = \frac{1}{4} + \eta, \quad \langle \Ss_{0,l}\Z_{2l-1}\Z_{2l}\rangle = \frac{1}{4} - \eta, \quad \langle \Ss_{0,l}\X_{2l-1}\X_{2l}\rangle = \frac{1}{4} - \eta, \quad \langle \Ss_{0,l}\Y_{2l-1}\Y_{2l}\rangle = -\frac{1}{4} + \eta.
\end{align*}
In this case the value of the first line fromn \eqref{papir} is equal to $2\eta$. By summing all the terms we obtain for \eqref{glavna}
\begin{align}\label{glavna1}
\bigg\| \Ss_{l,0}\ket{\psi} - \frac{\mathbb{I} + \Z_{2l-1}\Z_{2l} + \X_{2l-1}\X_{2l}- \Y_{2l-1}\Y_{2l}}{4}\ket{\psi} \bigg\| \leq O(\eta^{\frac{1}{2}} + \epsilon^m).
\end{align}
%
%
%
Similar robust versions of Eqs. (\ref{M1}-\ref{N3}) can be obtained, each having the same robustness bound.
Furthermore, using triangle inequality and relations analogous to \eqref{glavna1} the following bounds can be obtained:
\begin{align}
\label{ZzRob}
&\left\|\Z_{2l-1}^{(k)}\Z_{2l}^{(k)}\ket{\psi} - \left(\Ss_{l,0} + \Ss_{l,1} - \Ss_{l,2} - \Ss_{l,3}\right)\ket{\psi}\right\| = \\ \nonumber
& = \left\|\Z_{2l-1}^{(k)}\Z_{2l}^{(k)}\ket{\psi} - \Z_{2l-1}\Z_{2l}\ket{\psi} + \Z_{2l-1}\Z_{2l}\ket{\psi}  - \left(\Ss_{l,0} + \Ss_{l,1} - \Ss_{l,2} - \Ss_{l,3}\right)\ket{\psi}\right\|
\\ \nonumber
&\leq \left\|\Z_{2l-1}^{(k)}\Z_{2l}^{(k)}\ket{\psi} - \Z_{2l-1}\Z_{2l}\ket{\psi}\right\| + \left\|\Z_{2l-1}\Z_{2l}\ket{\psi}  - \left(\Ss_{l,0} + \Ss_{l,1} - \Ss_{l,2} - \Ss_{l,3}\right)\ket{\psi}\right\|
\\ \nonumber
&\leq O(\epsilon^m) + \Bigg\|\frac{\Z_{2l-1}\Z_{2l} + \mathbb{I} + \X_{2l-1}\X_{2l} - \Y_{2l-1}\Y_{2l}}{4}\ket{\psi} + \frac{\Z_{2l-1}\Z_{2l} + \mathbb{I} - \X_{2l-1}\X_{2l} + \Y_{2l-1}\Y_{2l}}{4}\ket{\psi} + \\ \nonumber &+ \frac{\Z_{2l-1}\Z_{2l} - \mathbb{I} + \X_{2l-1}\X_{2l} + \Y_{2l-1}\Y_{2l}}{4}\ket{\psi}+ \frac{\Z_{2l-1}\Z_{2l} - \mathbb{I} - \X_{2l-1}\X_{2l} - \Y_{2l-1}\Y_{2l}}{4}\ket{\psi}  - \left(\Ss_{l,0} + \Ss_{l,1} - \Ss_{l,2} - \Ss_{l,3}\right)\ket{\psi} \Bigg\| \\ \nonumber
&\leq O(\epsilon^m)+ \left\| \left(\Ss_{l,0} - \frac{\Z_{2l-1}\Z_{2l} + \mathbb{I} + \X_{2l-1}\X_{2l} - \Y_{2l-1}\Y_{2l}}{4} \right)\ket{\psi} \right\| + 
\left\| \left(\Ss_{l,1} - \frac{\Z_{2l-1}\Z_{2l} + \mathbb{I} - \X_{2l-1}\X_{2l} + \Y_{2l-1}\Y_{2l}}{4} \right)\ket{\psi} \right\| + \\ \nonumber
&+ \left\| \left(\Ss_{l,2} + \frac{\Z_{2l-1}\Z_{2l} - \mathbb{I} - \X_{2l-1}\X_{2l} - \Y_{2l-1}\Y_{2l}}{4} \right)\ket{\psi} \right\| + 
\left\| \left(\Ss_{l,3} + \frac{\Z_{2l-1}\Z_{2l} - \mathbb{I} + \X_{2l-1}\X_{2l} + \Y_{2l-1}\Y_{2l}}{4} \right)\ket{\psi} \right\|  \\ \nonumber
&\leq O(\eta^{\frac{1}{2}}+ \epsilon^m),
\end{align}
and similarly
\begin{align}\label{XxRob}
&\left\|\X_{2l-1}^{(k)}\X_{2l}^{(k)}\ket{\psi} - \left(\Ss_{l,0} - \Ss_{l,1} + \Ss_{l,2} - \Ss_{l,3}\right)\ket{\psi}\right\| \leq  O(\eta^{\frac{1}{2}}+ \epsilon^m), \\ \label{YyRob}
&\left\|\Y_{2l-1}^{(k)}\Y_{2l}^{(k)}\ket{\psi} - \left(-\Ss_{l,0} + \Ss_{l,1} + \Ss_{l,2} - \Ss_{l,3}\right)\ket{\psi}\right\|\leq O(\eta^{\frac{1}{2}}+ \epsilon^m).
\end{align}
The robust analogue of \eqref{XZY} is obtained through the following chain of inequalities
\begin{align}\label{XZYRob}
&\left\|\X_{2l-1}^{(k)}\X_{2l}^{(k)}\Z_{2l-1}^{(k)}\Z_{2l}^{(k)}\ket{\psi} + \Y_{2l-1}^{(k)}\Y_{2l}^{(k)}\ket{\psi} \right\| \\ \nonumber
= & \left\|\X_{2l-1}^{(k)}\X_{2l}^{(k)}\Z_{2l-1}^{(k)}\Z_{2l}^{(k)}\ket{\psi} - \X_{2l-1}^{(k)}\X_{2l}^{(k)}\left(\Ss_{l,0} + \Ss_{l,1} - \Ss_{l,2} - \Ss_{l,3}\right)\ket{\psi} + \X_{2l-1}^{(k)}\X_{2l}^{(k)}\left(\Ss_{l,0} + \Ss_{l,1} - \Ss_{l,2} - \Ss_{l,3}\right)\ket{\psi} + \Y_{2l-1}^{(k)}\Y_{2l}^{(k)}\ket{\psi} \right\| \\ \nonumber
\leq & O(\eta^{\frac{1}{2}}+ \epsilon^m) + \left\|\X_{2l-1}^{(k)}\X_{2l}^{(k)}\left(\Ss_{l,0} + \Ss_{l,1} - \Ss_{l,2} - \Ss_{l,3}\right)\ket{\psi} + \Y_{2l-1}^{(k)}\Y_{2l}^{(k)}\ket{\psi} \right\| \\ \nonumber
= & O(\eta^{\frac{1}{2}}+ \epsilon^m) + \big\|\X_{2l-1}^{(k)}\X_{2l}^{(k)}\left(\Ss_{l,0} + \Ss_{l,1} 
 - \Ss_{l,2} - \Ss_{l,3}\right)\ket{\psi}
-  \left(\Ss_{l,0} + \Ss_{l,1}  - \Ss_{l,2} - \Ss_{l,3}\right)\left(\Ss_{l,0} - \Ss_{l,1} + \Ss_{l,2} - \Ss_{l,3}\right)\ket{\psi}
+\\ \nonumber &\qquad \quad + \left(\Ss_{l,0} + \Ss_{l,1}  - \Ss_{l,2} - \Ss_{l,3}\right)\left(\Ss_{l,0} - \Ss_{l,1} + \Ss_{l,2} - \Ss_{l,3}\right)\ket{\psi}
  + \Y_{2l-1}^{(k)}\Y_{2l}^{(k)}\ket{\psi} \big\| \leq  O(\eta^{\frac{1}{2}}+ \epsilon^m).
\end{align}
To obtain the first inequality we used \eqref{ZzRob} and the fact that multiplication by a unitary ($\X_{2l-1}^{(k)}\X_{2l}^{(k)}$) does not change the norm. The last inequality is the consequence of \eqref{XxRob}, \eqref{YyRob} and the fact that $\Ss_{l,0} + \Ss_{l,1}  - \Ss_{l,2} - \Ss_{l,3}$ is a unitary operator. In a similar manner one can obtain
\begin{equation}\label{XZY+1Rob}
\left\|\X_{2l}^{(k)}\X_{2l+1}^{(k)}\Z_{2l}^{(k)}\Z_{2l+1}^{(k)}\ket{\psi} + \Y_{2l}^{(k)}\Y_{2l+1}^{(k)}\ket{\psi}\right\| \leq O(\eta^{\frac{1}{2}}+ \epsilon^m).
\end{equation}
Finally, to obtain \eqref{last} for $2^n - 2$ different values of $l$ one of two inequalities \eqref{XZYRob} and \eqref{XZY+1Rob} is used eight times (see \eqref{lancel}), thus leading to the final bound.
\begin{equation*}
\left\|\ket{\xi} - \ket{\xi_{0}}\tp\ket{0\dots 0} - \ket{\xi_{1}}\tp\ket{1\dots 1}\right\| \leq O(n^{\frac{1}{2}}(\eta^{\frac{1}{2}}+ \epsilon^m)).
\end{equation*}


\section{Entanglement certification proofs: qubits}
\subsection{Positivity of $\mathcal{I}$ for separable states: qubits}\label{entcertqubits}
Our aim is to prove that under maximal violation in step (ii) of the protocol
\begin{equation}
\label{ineq}
\mathcal{I}=\sum_{cduw}\omega_{cd}^{zw}\;p(c,+,+,d\vert z,x=\star,y=\star,w)\geq 0,
\end{equation}
holds for all separable $\varrho^{\rA\rB}$. First, note that the projectors for Charlie's measurement can be compactly written 
\begin{align}\label{projectors}
\Pi_{c|z}^{\rC'\rC''}=U_\rC^{\dagger}\sum_{j}\left(\pi_{c|z}^{\rC'}\right)^{T^{j}}\otimes\proj{j}^{\rC''}U_\rC,
\end{align}
where $U_\rC$ is the local unitary from lemma 1 and $\pi_{c|z}$ are projectors onto the Pauli eigenvectors, i.e. $\pi_{c|z}=\frac{1}{2}[\idd+c\sigma_{z}]$ for $\sigma_z=\sigma_{\tx{z}},\sigma_{\tx{x}},\sigma_{\tx{y}}$. Thus, at maximum violation, the (sub-normalized) states that Alice receives in the $\rA_0$ spaces conditional on a certain $c,z$ are given by
\begin{align}\label{steerqubits}
\tau_{c|z}=\frac{1}{2}\,U_\rA^{\dagger}\left[\sum_{j}\varrho_{\xi}^j\otimes(\pi_{c|z}^{\rA_0'})^{T_j}\right]U_\rA,
\end{align}
where 
\begin{align}
\varrho_{\xi}^j=\tr_{\rC''\rC\rC'}\left[\proj{j}^{\rC''}\,\proj{\xi}^{\rC''\rC\rA_0''\rA_0}\right].
\end{align}
Here, we have used the property $\tr_{\rC}[\proj{\Phi^{\tx{+}}} C\otimes\openone]=C^T$. We thus have
\begin{align}
p(c,+,+,d\vert z,x=\star,y=\star,w)&=\tr\left[\M_{+\vert\star}^{\rA_0\rA}\otimes\M_{+\vert\star}^{\rB_0\rB}\,\tau_{c|z}\otimes\varrho^{\rA\rB}\otimes\tau_{d|w}\right]\\
&=\sum_{j,k}\tr\left[\A\otimes\B\,\varrho_{\xi}^j\otimes(\pi_{c|z}^{\rA_0'})^{T_j}\otimes\varrho^{\rA\rB}\otimes(\pi_{d|w}^{\rB_0'})^{T_k}\otimes\varrho_{\xi}^k\right],
\end{align}
where $\A=\frac{1}{2}U_\rA\M_{+\vert\star}^{\rA_0\rA}U_\rA^\dagger$, $\B=\frac{1}{2}U_\rB\M_{+\vert\star}^{\rB\rB_0}U_\rB^\dagger$. Now, assume that $\varrho^{\rA\rB}$ is product so that $\varrho^{\rA\rB}=\sigma^\rA\otimes\sigma^\rB$ (mixtures of such states will be considered later). Then the above takes the form
\begin{align}
\sum_{j,k}\tr\left[ \pi_{c|z}^{T_j} \otimes \pi_{d|w}^{T_k} \, \A_j\otimes \B_k\right]
\end{align}
where 
\begin{align}
\A_j=\tr_{\rA\rA_0\rA_0''}\left[\A\, \varrho_{\xi}^j\otimes \idd_{\rA_0'} \otimes \sigma^{\rA}\right];\quad \B_k=\tr_{\rB\rB_0\rB_0''}\left[\B\,  \sigma^{\rB} \otimes \idd_{\rB_0'} \otimes \varrho_{\xi}^k\right].
\end{align}
Note that $\A_j$ and $\B_k$ are positive operators since $\A_j$ can be seen as a positive map applied to $\sigma^{\rA}$. Using this we may now write $\mathcal{I}$ as 
\begin{align}
\mathcal{I}&=\sum_{jk}\sum_{cdzw} \omega_{cd}^{zw}\tr\left[\pi_{c|z}^{T_j} \otimes \pi_{d|w}^{T_k} \, \A_j\otimes \B_k\right] \\
&=\sum_{jk}\sum_{cdzw} \omega_{cd}^{zw}\tr\left[\pi_{c|z} \otimes \pi_{d|w} \, \A_j^{T_j}\otimes \B_k^{T_k}\right] \\
&=\sum_{jk}\tr\left[\mathcal{W}\, \A_j^{T_j}\otimes \B_k^{T_k}\right]\geq 0,
\end{align}
where the second equality follows from $\tr[X]=\tr[X^T]$, and the final inequality follows from the fact that $\A_j^{T_j}$ and $\B_k^{T_k}$ are positive operators and thus $\A_j^{T_j}\otimes \B_k^{T_k}$ is a unnormalised product state. Since $\mathcal{I}$ is linear in $\varrho^{\rA\rB}$ one also has $\mathcal{I}\geq0$ for mixtures of product states and thus all separable states. 

\subsection{Positivity of $\mathcal{I}$ for separable states: arbitrary dimension}\label{entcertgen}
The proof follows the same structure as for the qubit case. As a consequence of Lemma 3, we have that Alice receives the subnormalised steered states conditioned on $\vz$, $\vc$: 
\begin{align}\label{steeredstatesgen}
\tau_{\vc,\vz}=\frac{1}{d}\,U_{\rA}^{\dagger}\left[\sum_{j=0}^{1}\varrho_{\xi}^j \otimes \left(\pi_{\vc|\vz}^{\rA_{0}'}\right)^{T^j}\right]U_{\rA},
\end{align}
where we define 
\begin{align}
\pi_{\vc|\vz}^{\rA_{0}'}=\otimes_i \,\pi_{c_i|z_i}^{\rA_{0i}'}\quad\quad \text{and} \quad\quad \varrho_{\xi}^j=\tr_{\rC''\rC\rC'}\left[(\otimes_i\proj{j}^{\rC''_i})\,\proj{\xi}^{\rC''\rC\rA_0''\rA_0}\right].
\end{align}
and
Bob has analogous states conditioned on Daisy's input and output. Now, the probabilities are given by
\begin{align}
p(\vc,+,+,\vd\vert \vz,x=\star,y=\star,\vw)&=\tr\left[\M_{+\vert\star}^{\rA_0\rA}\otimes\M_{+\vert\star}^{\rB_0\rB}\,\tau_{\vc|\vz}\otimes\varrho^{\rA\rB}\otimes\tau_{\vd|\vw}\right]\\
&=\sum_{j,k}\tr\left[\A\otimes\B\,\varrho_{\xi}^j\otimes(\pi_{\vc|\vz}^{\rA_0'})^{T_j}\otimes\varrho^{\rA\rB}\otimes(\pi_{\vd|\vw}^{\rB_0'})^{T_k}\otimes\varrho_{\xi}^k\right],
\end{align}
and $A=\frac{1}{d}U_\rA\M_{+\vert\star}^{\rA_0\rA}U_\rA^\dagger$, $\B=\frac{1}{d}U_\rB\M_{+\vert\star}^{\rB\rB_0}U_\rB^\dagger$. For separable $\varrho^{\rA\rB}=\sigma^{\rA}\otimes\sigma^{\rB}$ this takes the form
\begin{align}
p(\vc,+,+,\vd\vert \vz,x=\star,y=\star,\vw)=\sum_{j,k}\tr\left[ \pi_{\vc|\vz}^{T_j} \otimes \pi_{\vd|\vw}^{T_k} \, \A_j\otimes \B_k\right]
\end{align}
where again we have the positive operators
\begin{align}
\A_j=\tr_{\rA\rA_0\rA_0''}\left[\A\, \varrho_{\xi}^j\otimes \idd_{\rA_0'} \otimes \sigma^{\rA}\right];\quad \B_k=\tr_{\rB\rB_0\rB_0''}\left[\B\,  \sigma^{\rB} \otimes \idd_{\rB_0'} \otimes \varrho_{\xi}^k\right].
\end{align}
Hence we find
\begin{align}
\mathcal{I}&=\sum_{jk}\sum_{\vc\vd\vz\vw} \omega_{\vc\vd}^{\vz\vw}\tr\left[\pi_{\vc|\vz}^{T_j} \otimes \pi_{\vd|\vw}^{T_k} \, \A_j\otimes \B_k\right] \\
&=\sum_{jk}\sum_{\vc\vd\vz\vw} \omega_{\vc\vd}^{\vz\vw}\tr\left[\pi_{\vc|\vz} \otimes \pi_{\vd|\vw} \, \A_j^{T_j}\otimes \B_k^{T_k}\right] \\
&=\sum_{jk}\tr\left[\mathcal{W}\, \A_j^{T_j}\otimes \B_k^{T_k}\right]\geq 0.
\end{align}
Again, due to the linearity of $\mathcal{I}$ in $\varrho^{\rA\rB}$, one has $\mathcal{I}\geq 0$ for all separable states, completing the proof.

\section{Robust entanglement certification} \label{REC}

In this section we prove a relation \eqref{robustI} from the main text. We start from robust self-testing statements for Lemma 3
\begin{align}\label{viserysSupp}
\begin{split}
\bigg\| U[\lket{\psi}\otimes \lket{00}]&-\lket{\xi}\otimes \left[{\displaystyle \otimes_{i=1}^n}\lket{\Phi^+}^{\rC'_i\rA_i'}\right] \bigg\| \leq \theta \\
\bigg\| U\left[\Z_j^\rC\lket{\psi}\otimes\lket{00}\right]&- \lket{\xi} \otimes\left[ {\sigma_{\tx{z}}}^{\rC_j'} \tp_{i=1}^n{\lket{\Phi^{\tx{+}}}}^{\rC_i'\rA_i'}\right]\bigg\| \leq \theta,\\ 
\bigg\| U\left[\X_j^\rC\lket{\psi}\otimes\lket{00}\right]&- \lket{\xi}\otimes\left[{\sigma_{\tx{x}}}^{\rC_j'}\tp_{i=1}^n{\lket{\Phi^{\tx{+}}}}^{\rC_i'\rA_i'}\right]\bigg\| \leq \theta,\\ 
\bigg\| U\left[\Y_j^\rC\lket{\psi}\otimes\lket{00}\right]&- {\sigma_{\tx{z}}}^{\rC_j''}\lket{\xi}\otimes\left[{\sigma_{\tx{y}}}^{\rC_j'}
\tp_{i=1}^n{\lket{\Phi^{\tx{+}}}}^{\rC_i'\rA_i'}\right]\bigg\| \leq \theta,
\end{split}
\end{align}
and similarly for Daisy's measurements. These inequalities imply
\begin{align}\label{daenerys}
\begin{split}
U[\lket{\psi}\otimes \lket{00}] = \lket{\xi}\otimes \left[{\displaystyle \otimes_{i=1}^n}\lket{\Phi^+}^{\rC'_i\rA_i'}\right] + \lket{\hat{\Omega}},\\
U[\Z_j\lket{\psi}\otimes \lket{00}] = \lket{\xi}\otimes \left[\sigma_{\Z}^{\rC'_j}{\displaystyle \otimes_{i=1}^n}\lket{\Phi^+}^{\rC'_i\rA_i'}\right] + \lket{\hat{\Omega}_{\Z_j}},\\
U[\X_j\lket{\psi}\otimes \lket{00}] = \lket{\xi}\otimes \left[\sigma_{\X}^{\rC'_j}{\displaystyle \otimes_{i=1}^n}\lket{\Phi^+}^{\rC'_i\rA_i'}\right] + \lket{\hat{\Omega}_{\X_j}},\\
U[\Y_j\lket{\psi}\otimes \lket{00}] = \lket{\xi}\otimes \left[\sigma_{\Y}^{\rC'_j}{\displaystyle \otimes_{i=1}^n}\lket{\Phi^+}^{\rC'_i\rA_i'}\right] + \lket{\hat{\Omega}_{\Y_j}},
\end{split}
\end{align}
where $\ket{\hat{\Omega}},\ket{\hat{\Omega}_{\Z_j}}$, $\ket{\hat{\Omega}_{\X_j}}$ all have vector norm smaller than or equal to $\theta$. Let us concentrate on the first two equations from \eqref{daenerys} to get
\begin{equation}\label{robert}
U\left[\frac{\mathbb{I}\pm \Z_j}{2}\lket{\psi}\otimes \lket{00}\right] = \lket{\xi}\otimes \left[\frac{\mathbb{I}\pm \sigma_{\Z}^{\rC'_j}}{2} {\displaystyle \otimes_{i=1}^n}\lket{\Phi^+}^{\rC'_i\rA_i'}\right] + \lket{\Omega_{\Z_j}^{\pm}}
\end{equation}
where
\begin{align*}
\lket{\Omega_{\Z_j}^{\pm}} = \frac{1}{2}\left(\lket{\hat{\Omega}} \pm \lket{\hat{\Omega}_{\Z_j}}\right)
\end{align*}
is such that
\begin{equation}\label{selyse}
\left\|\lket{\Omega_{\Z_j}^{\pm}}\right\| \leq \frac{1}{2}\left(\left\|\lket{\hat{\Omega}_{\Z_j}}\right\| + \left\|\lket{\hat{\Omega}}\right\|\right) = \theta,
\end{equation}
due to the triangle inequality. Let us also recall that 
\begin{equation}\label{shireen}
\left\| \lket{\xi}\otimes \left[\frac{\mathbb{I}\pm \sigma_{\Z}^{\rC'_j}}{2} {\displaystyle \otimes_{i=1}^n}\lket{\Phi^+}^{\rC'_i\rA_i'}\right] \right\| = \left\| \lket{\psi_{Z,\pm}} \right\| = \frac{1}{\sqrt{2}}.
\end{equation}
The sub-normalised state Alice receives after Charlie measures $\Z_j$ and obtains $\pm 1$ is
\begin{eqnarray}\label{stanis}
\hat{\tau}_{\Z_j, \pm} &=& \tr_{C''CC'}\left[U_A^{\dagger} \left( \lketbra{\psi_{Z,\pm}}{\psi_{Z,\pm}} + \lketbra{\Omega_{\Z_j}^{\pm}}{\Omega_{\Z_j}^{\pm}} + \lketbra{\psi_{Z,\pm}}{\Omega_{\Z_j}^{\pm}} + \lketbra{\Omega_{\Z_j}^{\pm}}{\psi_{Z,\pm}} \right)U_A\right] \\ \nonumber
&=&  \tr_{C''CC'}\left[U_A^{\dagger} \left( \lketbra{\psi_{Z,\pm}}{\psi_{Z,\pm}} + \Delta_{\Z_j}^{\pm} \right)U_A\right].
\end{eqnarray}
It is useful to estimate trace norm $\|M\|_1=\tr\vert M\vert$ of operator $\Delta_{\Z_j}^{\pm}$. For that purpose we use triangle inequality
\begin{eqnarray}\nonumber
\left\| \Delta_{\Z_j}^{\pm} \right \|_1 \  &=& \left\| \lketbra{\Omega_{\Z_j}^{\pm}}{\Omega_{\Z_j}^{\pm}} + \lketbra{\psi_{Z,\pm}}{\Omega_{\Z_j}^{\pm}} + \lketbra{\Omega_{\Z_j}^{\pm}}{\psi_{Z,\pm}}\right \|_1 \\ \label{jofrey}
&\leq &  \left\| \lketbra{\Omega_{\Z_j}^{\pm}}{\Omega_{\Z_j}^{\pm}} \right \|_1 + \left\|\lketbra{\psi_{Z,\pm}}{\Omega_{\Z_j}^{\pm}}\right \|_1 + \left\|\lketbra{\Omega_{\Z_j}^{\pm}}{\psi_{Z,\pm}}\right. \|_1 \\ \nonumber
\end{eqnarray}
Let us now estimate the trace norm of each term separately, starting from the first term
\begin{eqnarray*}
\left\| \lketbra{\Omega_{\Z_j}^{\pm}}{\Omega_{\Z_j}^{\pm}} \right \|_1 = \tr\left(\left|\lketbra{\Omega_{\Z_j}^{\pm}}{\Omega_{\Z_j}^{\pm}}\right|\right)= 
\tr\left(\lketbra{\Omega_{\Z_j}^{\pm}}{\Omega_{\Z_j}^{\pm}}\right) = \tr\left(\lbraket{\Omega_{\Z_j}^{\pm}}{\Omega_{\Z_j}^{\pm}}\right) \leq \theta^2.
\end{eqnarray*}
The first equality is just the definition of the trace norm, the second uses positivity of $\ketbra{\Omega_{\Z_j}^{\pm}}{\Omega_{\Z_j}^{\pm}}$, and the inequality follows from \eqref{selyse}. Trace norm of the second term from \eqref{jofrey} can be bounded in the following way
\begin{equation*}
\left\|\lketbra{\psi_{Z,\pm}}{\Omega_{\Z_j}^{\pm}}\right \|_1 = \tr\left(\sqrt{\lketbra{\psi_{Z,\pm}}{\Omega_{\Z_j}^{\pm}}\lketbra{\Omega_{\Z_j}^{\pm}}{\psi_{Z,\pm}}}\right) \leq \frac{\theta}{\sqrt{2}} 
\end{equation*}
where the inequality follows from \eqref{selyse} and norm of $\lket{\psi_{Z,\pm}}$. Finally the trace norm of third term from \eqref{jofrey} is
\begin{equation*}
\left\|\lketbra{\Omega_{\Z_j}^{\pm}}{\psi_{Z,\pm}}\right \|_1 = \tr\left(\sqrt{\lketbra{\Omega_{\Z_j}^{\pm}}{\psi_{Z,\pm}}\lketbra{\psi_{Z,\pm}}{\Omega_{\Z_j}^{\pm}}}\right) = \frac{1}{\sqrt{2}}\tr\left(\sqrt{\lketbra{{\Omega_{\Z_j}^{\pm}}}{{\Omega_{\Z_j}^{\pm}}}}\right) \leq \frac{\theta}{\sqrt{2}}.
\end{equation*}
To get the last inequality we used the relation
\begin{equation*}
\tr\left(\sqrt{\lketbra{{\Omega_{\Z_j}^{\pm}}}{{\Omega_{\Z_j}^{\pm}}}}\right) = \tr\left(\sqrt{\lbraket{{\Omega_{\Z_j}^{\pm}}}{{\Omega_{\Z_j}^{\pm}}}\frac{\lketbra{{\Omega_{\Z_j}^{\pm}}}{{\Omega_{\Z_j}^{\pm}}}}{\lbraket{{\Omega_{\Z_j}^{\pm}}}{{\Omega_{\Z_j}^{\pm}}}}}\right) \leq \theta
\end{equation*}
where the last inequality comes from the fact that ${\lketbra{{\Omega_{\Z_j}^{\pm}}}{{\Omega_{\Z_j}^{\pm}}}}/{\lbraket{{\Omega_{\Z_j}^{\pm}}}{{\Omega_{\Z_j}^{\pm}}}}$ 
is a projector. Finally, \eqref{jofrey} reduces to
\begin{equation}\label{tommen}
\left\| \Delta_{\Z_j}^{\pm} \right \|_1 \leq \sqrt{2}\theta + \theta^2
\end{equation}
An equivalent bound can be obtained when Charlie measures $\X_j$ or $\Y_j$.  
By rewriting \eqref{jofrey}, we can see that Alice's steered states have the following form
\begin{eqnarray*}
\hat{\tau}_{\vc |\vz } &=&  \tau_{\vc |\vz } + \Delta_{\vc | \vz}, \qquad \forall \vc, \vz
\end{eqnarray*}
where $\tau_{\vc |\vz }$ are the ideal steered states given in \eqref{steeredstatesgen}. Depending on $\vc$ and $\vz$ the operators $\Delta_{\vc | \vz}$ are obtained by tracing out Charlie's system from the corresponding $\Delta_{\textrm{P}_j}^{\pm}$, with $\textrm{P} \in \{\Z,\X,\Y\}$. For every $\vc$ and $\vz$ the correction states $\Delta_{\vc | \vz}$ have bounded trace norm
\begin{equation}\label{olena}
\left\| \Delta_{\vc | \vz} \right \|_1 = \left\| \tr_{CC'C''}\left(\Delta_{\textrm{P}_j}^{\pm}\right) \right \|_1  \leq \left\| \Delta_{\textrm{P}_j}^{\pm} \right \|_1  \leq \sqrt{2}\theta + \theta^2
\end{equation}
The first inequality comes from the fact that trace norm cannot increase by performing partial trace \cite{Rastegin}.
Similarly, Bob's steered states have form
\begin{align}\label{mace}
\begin{split}
& \hat{\tau}_{\vd |\vw } =  \tau_{\vd |\vw } + \Delta_{\vd | \vw}, \\
& \left\| \Delta_{\vd | \vw} \right \|_1 \leq  \sqrt{2}\theta + \theta^2.
\end{split}
\end{align}
Equiped with characterization of Alice's and Bob's steered states let us estimate the lowest value of $\mathcal{I}$ from \eqref{jaime} when evaluated on a separable state $\varrho^{AB} = \sum_{\lambda}p_{\lambda}\varrho^A_{\lambda}\otimes \varrho^B_{\lambda}$:
\begin{align}\label{margeory}
\begin{split}
\mathcal{I} &= \sum_{\lambda}p_{\lambda}\sum_{\vc,\vd,\vz,\vw}\omega_{\vc,\vd}^{\vz,\vw}\tr\left[\M_{+\vert\star}^{\rA_0\rA}\otimes\M_{+\vert\star}^{\rB_0\rB}\,\hat{\tau}_{\vc|\vz}\otimes\varrho^{\rA}_{\lambda}\otimes\varrho^{\rB}_{\lambda}\otimes\hat{\tau}_{\vd|\vw}\right] \\ 
&= \sum_{\lambda}p_{\lambda}\sum_{\vc,\vd,\vz,\vw}\omega_{\vc,\vd}^{\vz,\vw}\tr\left[\M_{+\vert\star}^{\rA_0\rA}\otimes\M_{+\vert\star}^{\rB_0\rB}\,\left(\tau_{\vc|\vz}+\Delta_{\vc | \vz}\right)\otimes\varrho^{\rA}_{\lambda}\otimes\varrho^{\rB}_{\lambda}\otimes\left(\tau_{\vd|\vw}+\Delta_{\vd | \vw}\right)\right] \\
&= \mathcal{I}_{\textrm{noiseless}} + \sum_{\lambda}p_{\lambda}\sum_{\vc,\vd,\vz,\vw}\omega_{\vc,\vd}^{\vz,\vw}\bigg\{
\tr\left[\M_{+\vert\star}^{\rA_0\rA} \Delta_{\vc | \vz} \otimes\varrho^{\rA}_{\lambda} \right]
\tr\left[\M_{+\vert\star}^{\rB_0\rB}\tau_{\vd |\vw} \otimes\varrho^{\rB}_{\lambda}\right] +  \\ &\qquad +  \tr\left[\M_{+\vert\star}^{\rA_0\rA} \tau_{\vc|\vz} \otimes\varrho^{\rA}_{\lambda}\right]
\tr\left[\M_{+\vert\star}^{\rB_0\rB}\Delta_{\vd|\vw} \otimes\varrho^{\rB}_{\lambda}\right] + \tr\left[\M_{+\vert\star}^{\rA_0\rA} \Delta_{\vc | \vz} \otimes\varrho^{\rA}_{\lambda}\right]
\tr\left[\M_{+\vert\star}^{\rB_0\rB}\Delta_{\vd|\vw} \otimes\varrho^{\rB}_{\lambda}\right] 
\bigg\}.
\end{split}
\end{align}
$\mathcal{I}_{\textrm{noiseless}}$ is the value $\mathcal{I}$ would have in the ideal case $\theta = 0$. To estimate how negative the total value of $\mathcal{I}$ given in \eqref{margeory} can be, we assume the worst case, i.e. $\mathcal{I}_{\textrm{noiseless}} = 0$ and all other contributions give negative contribution. To bound the absolute value of those contributions note that
\begin{align}\label{wilas}
\begin{split}
\left|\tr\left[\M_{+\vert\star}^{\rA_0\rA} \Delta_{\vc | \vz} \otimes\varrho^{\rA}_{\lambda} \right]\right| &\leq \tr\left|\M_{+\vert\star}^{\rA_0\rA} \Delta_{\vc | \vz} \otimes\varrho^{\rA}_{\lambda} \right| \\ 
&= \left\| \M_{+\vert\star}^{\rA_0\rA} \Delta_{\vc | \vz} \otimes\varrho^{\rA}_{\lambda} \right\|_1 \\
&\leq \left\| \M_{+\vert\star}^{\rA_0\rA}\right\|_{\infty} \left\|\Delta_{\vc | \vz} \otimes\varrho^{\rA}_{\lambda} \right\|_1 \\
&\leq \tr\left(\left|\Delta_{\vc | \vz}\right|\right)\tr\left(\varrho^{\rA}_{\lambda}\right) \\
&= \left\|\Delta_{\vc | \vz}\right\|_1 \leq \sqrt{2}\theta + \theta^2.
\end{split}
\end{align}
The first line follows from the inequality $|\tr(A)|\leq \tr|A|$. To obtain the third line we used H\"{o}lder's inequality $\tr(AB) \leq \|A\|_\infty\|B\|_1$ \cite{holder1,holder2}.  The fourth lines uses the fact that infinite Schatten norm of $\M_{+\vert\star}^{\rA_0\rA}$ is its maximal eigenvalue which cannot be larger than one. Finally in the fifth line we used the fact that $\varrho^{\rA}_{\lambda}$ is a normalized state and  \eqref{olena}. By using the same argumentation one can show that
\begin{align}\label{garlan}
\left|\tr\left[\M_{+\vert\star}^{\rA_0\rA} \tau_{\vd | \vw} \otimes\varrho^{\rA}_{\lambda} \right]\right| \leq \frac{1}{2}.
\end{align}
If we plug \eqref{wilas}, \eqref{garlan} and their analogues obtained by transforming $(\vz,\vc)\leftrightarrow (\vw,\vd)$ into \eqref{margeory} we have that in the worst case
\begin{align*}
\mathcal{I} \sim O(\theta).
\end{align*}
%


\section{Entanglement certification of two-qubit Werner states}\label{isonoise}
\begin{figure}
\includegraphics[scale=0.7]{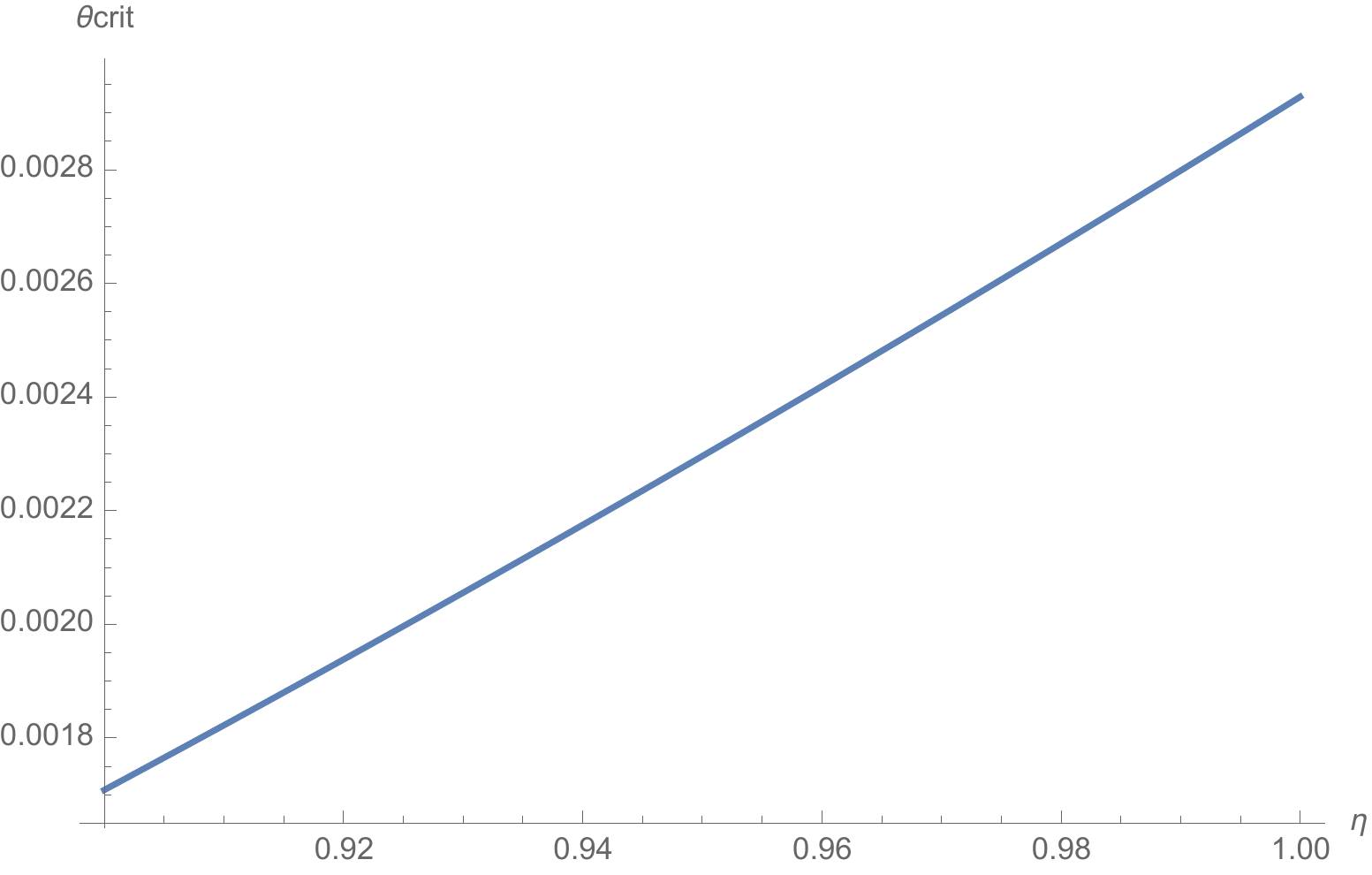}
\caption{\label{werner_fig} Critical robustness of self-testing needed to certify the entanglement of the Werner state with noise parameter 0.6 as a function of the visibility of the auxiliary states.}
\end{figure}
Here we analyse the effect of noise in the auxiliary states when certifying the entanglement of the two-qubit Werner states
\begin{align}\label{isostate}
\varrho_W(p)=p\proj{\Phi^{\text{+}}}+(1-p)\frac{\openone}{4},
\end{align} 
where $\ket{\Phi^{\text{+}}}=\frac{1}{\sqrt{2}}(\ket{00}+\ket{11})$. Note that the optimal entanglement witness for these state under white noise is 
\begin{align}\label{isowit}
W=\sigma_z\otimes\sigma_z+\sigma_y\otimes\sigma_y+\sigma_x\otimes\sigma_x+\mathbb{I}
\end{align}
One has $\text{Tr}[W\rho_{\text{SEP}}]\geq0$ and $\text{Tr}[W\varrho_{\text{W}}(p)]=1-3p$. From equations \eqref{margeory} - \eqref{garlan} and taking the worst case, one can certify entanglement using the above witness if 
\begin{align}
\mathcal{I}<-12[(\sqrt{2}\theta+\theta^2)^2+\sqrt{2}\theta+\theta^2]
\end{align}
where $\theta$ quantifies the robustness of self-testing (see \eqref{viserysSupp}). Here the number 12 comes from the number of terms in the decomposition of the witness \eqref{isowit} into products of Pauli projectors. Let us assume that the auxiliary states are also Werner states with visibility $\eta$. Assuming noiseless measurements, one would expect to observe a value
\begin{align}
\mathcal{I}=\frac{1}{16}((1-3p)\eta^2+2\eta(1-\eta)+(1-\eta)^2\frac{1}{4}),
\end{align}
since there is probability $\eta^2$  that the auxiliary states both produce a maximally entangled state and if one or no maximally entangled states are produced in the auxiliary states the value of the inequality will be 1/16 or 1/64 respectively. Thus, one is able to certify entanglement if
\begin{align}\label{wernercon}
\mathcal{I}=\frac{1}{16}((1-3p)\eta^2+2\eta(1-\eta)+(1-\eta)^2\frac{1}{4})<-12[(\sqrt{2}\theta+\theta^2)^2+\sqrt{2}\theta+\theta^2].
\end{align}
This inequality gives the condition that needs to be satisfied in order to be able to certify the entanglement of the state \eqref{isostate}. Note that $\theta$ will implicitly depend on $\eta$ through some robust self-testing statement. Given a particular $\eta$, one therefore needs to ensure that there is a robust self-testing statement with corresponding $\theta$ smaller than some critical $\theta_{\text{crit}}$ given by \eqref{wernercon}. In Fig.\ \ref{werner_fig} we plot the values of $\theta_{\text{crit}}$ for different values of $\eta$ and taking $p=0.6$ (note that the state has a local hidden variable model in the standard Bell scenario for this visibility \cite{grot,grot2}). 
For $\eta=1$ we have $\theta=0$ which is below $\theta_{\text{crit}}$. As one decreases $\eta$, at some point the $\theta$ given by the robust self-testing statement will be above the critical value and the method will not work. The question is then for which value of $\eta$ does this happen? Given the small values of $\theta_{\text{crit}}$ this will likely happen for a value of $\eta$ very close to 1. We do not go further into the analysis here; to get precise numbers one could use the methods we present in appendix \ref{lemma1Rob} or for better results try to extend the method in \cite{Kaniewski_ap} to the self-testing of measurements. We note however that very high visibilities can be achieved experimentally, e.g. using photonic set-ups visibilities of above 0.999 \cite{Poh} and 0.997 \cite{Gomez2018} have been reported.

\end{appendix}

\end{document}